\documentclass[lettersize,journal]{IEEEtran}

\usepackage{amssymb}
\usepackage{amsmath}
\usepackage{xcolor}
\usepackage{graphicx, epsfig, amssymb, amsmath, amsthm, bm}
\usepackage{multirow}
\usepackage{pict2e}
\usepackage{times}
\usepackage{cases}
\usepackage{array}
\usepackage{float}
\usepackage{algorithmic}
\usepackage[ruled]{algorithm2e}
   
\usepackage{setspace}
\usepackage{flushend}
\usepackage{stfloats}
\usepackage{amsmath}
\usepackage{subfigure}
\usepackage{booktabs}
\usepackage{multicol}

\newtheorem{theorem}{Theorem}

\newtheorem{lemma}{Lemma}

\newtheorem{definition}{Definition}

\allowdisplaybreaks

\begin{document}


\title{Low-Complexity Hybrid Beamforming for Multi-Cell mmWave Massive MIMO: \\ A Primitive Kronecker Decomposition Approach}

\author{Teng Sun, Guangxu Zhu, Xiaofan Li, Jiancun Fan, and  Minghua Xia
	\thanks{This work was supported by the Guangdong Basic and Applied Basic Research Project under Grant 2022A1515140166 and in part by the National Natural Science Foundation of China under Grant 62271232. {\it (Corresponding author: Minghua Xia)}}
	\thanks{Teng Sun is with the School of Electronic and Information Engineering, Xi'an Jiaotong University, Xi'an 710049, China, and also with the 54th Research Institute of China Electronics Technology Group Corporation, Shijiazhuang 050081, China (e-mail: 15176948810@126.com.)}
	\thanks{Guangxu Zhu is with Shenzhen Research Institute of Big Data, Shenzhen 710049, China (email: gxzhu@sribd.cn)}
	\thanks{Xiaofan Li is with the School of Intelligent Systems Science and Engineering, Jinan University, Zhuhai 519070, China (email: lixiaofan@jnu.edu.cn)}
	\thanks{Jiancun Fan is with the School of Electronic and Information Engineering, Xi'an Jiaotong University, Xi'an 710049, China (email: fanjc0114@gmail.com)}
	\thanks{Minghua Xia is with the School of Electronics and Information Technology, Sun Yat-sen University, Guangzhou 510006, China (e-mail: xiamingh@mail.sysu.edu.cn).} 
	}

\maketitle

 \maketitle
         
\begin{abstract}
To circumvent the high path loss of mmWave propagation and reduce the hardware cost of massive multiple-input multiple-output antenna systems, full-dimensional hybrid beamforming is critical in 5G and beyond wireless communications. Concerning an uplink multi-cell system with a large-scale uniform planar antenna array, this paper designs an efficient hybrid beamformer using primitive Kronecker decomposition and dynamic factor allocation, where the analog beamformer applies to null the inter-cell interference and simultaneously enhances the desired signals. In contrast, the digital beamformer mitigates the intra-cell interference using the minimum mean square error (MMSE) criterion. Then, due to the low accuracy of phase shifters inherent in the analog beamformer, a low-complexity hybrid beamformer is developed to slow its adjustment speed. Next, an optimality analysis from a subspace perspective is performed, and a sufficient condition for optimal antenna configuration is established. Finally, simulation results demonstrate that the achievable sum rate of the proposed beamformer approaches that of the optimal pure digital MMSE scheme, yet with much lower computational complexity and hardware cost. 
\end{abstract}



\section{Introduction} 
\label{Section_Introduction}
Integrating massive multiple-input multiple-output (MIMO) and mmWave techniques is a prospective candidate for next-generation wireless systems with ultra-high data rates \cite{8626085}. A hybrid digital and analog beamforming architecture was proposed to facilitate the implementation of massive MIMO systems to balance the hardware cost and beamforming gain. A hybrid beamformer consists of a small-scale digital and a large-scale analog beamformer. The former is implemented by a baseband signal processor cascaded with several RF chains in real-world applications. In contrast, the latter is realized by a phase shifter network. Due to the unit-modulus characteristics of the phase shifter network, the design of the hybrid beamformer is mathematically intractable, so some approximations can be applied in point-to-point or multi-user scenarios. As cell size shrinks in mmWave scenarios, the effect of inter-cell interference becomes dominant. Consequently, this work focuses on the realistic multi-cell case where intra-cell and inter-cell interferences are mitigated.

When applying a pure digital beamformer for single-user MIMO communications, the Shannon capacity can be achieved by performing singular value decomposition (SVD) of the underlying channel matrix \cite{1203154}. In contrast, when using a hybrid beamformer, the beamforming matrix can be constructed as the product of a digital beamforming matrix and an analog one, with the constraint that each entry must have the unit modulus. The problem can be intuitively framed as minimizing the Euclidean distance between the optimal pure digital and hybrid beamformers to achieve the Shannon capacity. Using the spatial structure of mmWave channels, an orthogonal matching pursuit scheme was designed in \cite{Ayach2014} to find the best low-dimensional approximation of the optimal pure digital beamformer. However, the performance of the orthogonal matching pursuit depends heavily upon the codebook size. In another approach, the study \cite{Yu2016} applied the manifold learning theory to minimize the Euclidean distance by optimizing the analog beamformer on a Riemannian manifold. Furthermore, the work \cite{Jin2018} relaxed the unit-modulus constraint and redefined the hybrid beamforming problem as an unconstrained optimization problem, which helps reduce both the Euclidean distance and computational complexity. More recently, a deep neural network was used in \cite{Peken2020} to approximate the SVD operation for hybrid beamforming. 

For multi-user MIMO scenarios in the presence of intra-cell interference, the optimal pure digital beamformer is the minimum mean square error (MMSE) scheme, which is information lossless but has very high computational complexity \cite[Sec. 8.3.4]{Tse2005}. Like the preceding single-user case, an intuitive idea for designing a hybrid beamformer for multi-user MIMO is to minimize the distance between the optimal MMSE beamformer and the hybrid one. For instance, some alternating iteration optimization methods were employed in \cite{Li2017, Nguyen2017}. On the other hand, the hybrid beamforming problem can be directly solved from the information theory perspective instead of minimizing the Euclidean distance. For instance, the joint design of digital and analog beamformers was performed using the penalty dual decomposition method in \cite{Shi2018}. However, the iterative process for the joint design has low convergence speed and/or high computational complexity. To deal with such difficulties, decoupling the digital and analog ones prevails in the context of massive MIMO \cite{Alkhateeb2015, Zhang2020, 9419762}, where the analog one is designed to maximize the received power. In contrast, the digital one suppresses multi-user interference. Recently, the work \cite{10355874} designed a group-and-codebook-based two-timescale hybrid precoding for reducing complexity by clustering users into groups based on their statistical similarity and selecting analog beams from an orthogonal codebook.

For massive multi-cell MIMO networks, inter-cell interference further degrades network performance if not properly handled, in addition to intra-cell multi-user interference. Even worse, in forthcoming 6G wireless networks where unmanned aerial vehicles (UAVs) serving as aerial access points (APs) are deemed to become popular, the inter-cell interference will be boosted because of the strong line-of-sight transmission between aerial APs and terrestrial BSs \cite{Liu2019, 10381632}. To mitigate inter-cell interference in hybrid beamforming design, a simple approach is to use a multi-user scheduling strategy, e.g., various coordinated multipoint (CoMP) techniques \cite{8976426}. In \cite{Nasir2020}, the inter-cell interference was suppressed by alternatively serving the cell-center and cell-edge users. In \cite{9349195}, a successive interference cancellation (SIC) strategy was applied to suppress inter-cell interference in the digital domain. By using the angles of arrival (AoAs) of interference channels at a {\it linear antenna array}, the work \cite{Zhu2017} proposed to null the inter-cell interference by an analog beamformer. In contrast, a digital beamformer mitigates the intra-cell interference. When antenna size grows more extensive, however, {\it full-dimensional MIMO} (FD-MIMO, also widely known as 3D MIMO), instead of the linear antenna array, is more suitable for real-world applications \cite{Xu2017}. In practice, FD-MIMO is realized by a two-dimensional active planar antenna array, which can significantly increase the spatial degrees of freedom (DoFs) by adjusting the elevation and azimuth angles to use three-dimensional channels fully \cite{7060514}.

Building on our previous work \cite{Zhu2017}, which focused on linear antenna arrays, a single-path channel model for inter-cell interference, and single-antenna UEs, this study addresses more practical FD-MIMO scenarios characterized by multi-path inter-cell interference and multi-antenna UEs. Compared to \cite{Zhu2017}, we introduce a primitive Kronecker decomposition approach that allows for optimal allocation of Kronecker factors to eliminate inter-cell interference and enhance the desired signals of intra-cell UEs. Moreover, considering the engineering fact that analog beamformers are typically implemented using phase shifters with lower accuracy than digital beamformers, simultaneous adjustment of both analog and digital beamformers is unnecessary. Consequently, we further develop a low-complexity hybrid beamformer, where the analog beamformer adapts much slower than the digital beamformer. Next, we conduct an optimality analysis from a subspace perspective and establish sufficient conditions for antenna configuration to implement the proposed hybrid beamformer. Simulation results indicate that the sum rate of the proposed hybrid beamformer approaches that of the optimal digital MMSE scheme while achieving significantly lower computational complexity and hardware costs. Additionally, the proposed hybrid beamformer outperforms benchmark schemes in terms of the achievable sum rate and adjustment speed of the analog beamformer. Both proposed schemes demonstrate robustness against varying inter-cell interference levels, thanks to the orthogonality of Kronecker factors and interference channels.

The remainder of this paper is organized as follows. Section~\ref{Section-PKD} defines the primitive Kronecker decomposition. Section~\ref{Section_SystemModel} introduces the system and mmWave channel models. Section~\ref{Section_ProblemFormulation} formulates the problem of hybrid beamformer design for maximizing the sum rate, while Section~\ref{Section_KroneckerBeamforming} elaborates on the design procedure for the hybrid beamformer. Next, Section~\ref{Section_OptimalityComputationalComplexity} conducts an optimality analysis from a subspace perspective and presents a sufficient condition of antenna configuration needed to implement the designed beamformer, along with an analysis of computational complexity. Section~\ref{Section_SimulationResults} presents and discusses simulation results, and finally, Section~\ref{Section_Conclusion} concludes the paper.
	
{\it Notation}: Vectors and matrices are denoted by lower- and upper-case letters in bold typeface, respectively. The operators $(\cdot)^T$, $(\cdot)^H$, and $\otimes$ denote transpose, conjugate transpose, and Kronecker product, respectively. The symbol $\emptyset$ denotes an empty set; $\bm{0}$ indicates a vector with all elements being zero; $\bm{I}_n$ stands for the $n \times n$ identity matrix; $\bm{E}_{ij, m \times n}$ is the $m \times n$ elementary matrix with the $(i, j)$-th entry being unity while the other entries being zero, and $\mathbb{C}^{m \times n}$ refers to the complex space of dimension $m \times n$. The ceiling operator $\lceil x \rceil$ maps $x$ to the least integer greater than or equal to $x$; the function ${\rm angle}(\bm{x})$ returns the vector with each element being the angle of the corresponding component of $\bm{x}$; ${\rm length}(\bm{x})$ gives the size of $\bm{x}$; ${\rm diag}(\bm{x})$ returns a diagonal matrix with diagonal entries specified by $\bm{x}$; $\|\bm{x}\|$ indicates the $\ell_2$-norm of $\bm{x}$, and ${\rm rank}(\bm{A})$ gives the rank of $\bm{A}$. Finally, the abbreviation $\mathcal{CN}(m,\sigma^2)$ refers to the circularly symmetric complex Gaussian distribution with mean $m$ and variance $\sigma^2$.

\section{Mathematical Preliminary: Primitive Kronecker Decomposition}
\label{Section-PKD}

In the theory of matrix analysis, the Kronecker decomposition of a vector with unit-modulus elements is defined in the following lemma \cite{Zhu2017}.
\begin{lemma}[Kronecker Decomposition]
	\label{lemma_KronDecom}
	Let $\bm{a} \triangleq \left[1, e^{j \Theta}, e^{j 2 \Theta}, \cdots, e^{j(N-1) \Theta}\right]^{T} \in \mathbb{C}^{N\times 1}$, where $\Theta$ is fixed. Suppose that $N = n_{1} \times n_{2} \times \cdots \times n_{D}$ with $\left\{n_{d}\right\}_{d=1}^{D}$ being positive integers; then the vector $\bm{a}$ can be decomposed into
	\begin{equation} \label{Eq-10}
		\bm{a} = \bm{a}^{(1)} \otimes \bm{a}^{(2)} \otimes \cdots \otimes \bm{a}^{(D)},
	\end{equation}
	where $\bm{a}^{(d)} = \left[1, e^{j \Omega}, e^{j 2 \Omega}, \cdots, e^{j\left(n_{d}-1\right) \Omega }\right]^T \in \mathbb{C}^{n_d \times 1}$, with $\Omega = n_{0} \times n_{1} \times \cdots \times n_{d-1} \times \Theta$ and $n_{0} \equiv 1$.
\end{lemma}

The key property of Lemma~\ref{lemma_KronDecom} is that both the vector $\bm{a}$ and its factors $\bm{a}^{(d)}$ for all $d = 1, \cdots, D$ are all phase-shift vectors with unit-modulus elements. To fully explore the potential of Kronecker decomposition, we seek to maximize the number of Kronecker factors, denoted as $D$ in \eqref{Eq-10}. Accordingly, we define the primitive Kronecker decomposition as follows.
\begin{definition}[Primitive Kronecker Decomposition]
\label{Def-PKD}
	Given $N = n_{1} \times n_{2} \times \cdots \times n_{D^{\prime}}$, where $\left\{n_{d}\right\}_{d=1}^{D^{\prime}}$ are the prime factors of $N$, the primitive Kronecker decomposition of $\bm{a}$ is defined as
	\begin{equation} \label{Eq-10-b}
		\bm{a} = \bm{a}^{(1)} \otimes \bm{a}^{(2)} \otimes \cdots \otimes \bm{a}^{(D^{\prime})}.
	\end{equation}
\end{definition}

The result \eqref{Eq-10-b} is always unique up to the order of the prime factors, based on the fundamental theorem of arithmetic \cite[Thm 2]{Hardy1938}. When compared to \eqref{Eq-10}, the primitive Kronecker decomposition presented in \eqref{Eq-10-b} offers two main advantages: {\it i)} it allows the shortest factor to reset $\bm{a}$ to $\bm{0}$ after performing the factor decomposition, and {\it ii)}  it maintains the maximum number of factors.

For comparison, let us consider $\bm{a} \in \mathbb{C}^{N \times 1}$ with $N = 12$. Utilizing simple integer factorization, we have various factorizations $N = 3 \times 4 = 2 \times 6 = 2 \times 2 \times 3$. The last factorization corresponds to the prime factorization, in which every factor is prime. Specifically, we can express $\bm{a}$ in the following ways:
\begin{align}
	\bm{a} &= \bm{a}_{1}^{(1)} \otimes \bm{a}_{1}^{(2)}, \label{Eq-10-b-1} \\
	\bm{a} &= \bm{a}_{2}^{(1)} \otimes \bm{a}_{2}^{(2)}, \label{Eq-10-b-2} \\
	\bm{a} &= \bm{a}_{3}^{(1)} \otimes \bm{a}_{3}^{(2)} \otimes \bm{a}_{3}^{(3)}, \label{Eq-10-b-3}
\end{align}
where $\bm{a}_{1}^{(1)} \in \mathbb{C}^{3 \times 1}$, $\bm{a}_{1}^{(2)} \in \mathbb{C}^{4 \times 1}$; $\bm{a}_{2}^{(1)} \in \mathbb{C}^{2 \times 1}$, $\bm{a}_{2}^{(2)} \in \mathbb{C}^{6 \times 1}$; and $\bm{a}_{3}^{(1)} \in \mathbb{C}^{2 \times 1}$, $\bm{a}_{3}^{(2)} \in \mathbb{C}^{2 \times 1}$, and $\bm{a}_{3}^{(3)} \in \mathbb{C}^{3 \times 1}$. In the case of \eqref{Eq-10-b-1}, we can set either $\bm{a}_{1}^{(1)} = \bm{0}$ or $\bm{a}_{1}^{(2)} = \bm{0}$, which will reset $\bm{a}$ to $\bm{0}$. Here, the shortest factor is $\bm{a}_{1}^{(1)}$, with a length of $3$. In contrast, the primitive Kronecker decomposition represented by \eqref{Eq-10-b-3} indicates that the minimal factor length is $2$. Although \eqref{Eq-10-b-2} shares the same shortest factor as \eqref{Eq-10-b-3}, the latter contains the maximum number of factors.

In engineering applications, such as the beamforming design discussed in this paper, the primitive Kronecker decomposition is essential. It enables us to minimize the DoFs required to null a vector while maximizing the remaining degrees. Moreover, incorporating more Kronecker factors enhances the ability to eliminate inter-cell interference from a subspace perspective. We will utilize these properties in our design later on.

\section{System and Channel Models}\label{Section_SystemModel}
This section begins with an uplink multi-cell FD-MIMO communication system model and then transitions to the mmWave channel model.

\subsection{Multi-Cell FD-MIMO System Model}
We consider the uplink transmission of a multi-cell mmWave FD-MIMO system. In the target cell, $K$ UEs communicate simultaneously with the BS at the cell center. Each UE is equipped with a uniform linear array (ULA) of $Q$ antenna elements and transmits a single data stream to the BS. The received signal at the BS is corrupted by $\Psi$ interferences from adjacent cells, where the $\psi^{\rm th}$ interference is assumed to have $\Gamma_\psi$ propagation paths, for all $\psi = 1, \cdots, \Psi$. Suppose that the BS is equipped with a UPA of size $M \times N$ where $M$ and $N$ denote the number of antenna rows and columns, respectively; its received signal can be expressed as
\begin{equation} \label{Eq-1}
	\bm{y} = \bm{GVx}+\bm{Hs}+\bm{z},
\end{equation} 
where $\bm{G} \triangleq [\bm{G}_1, \bm{G}_2, \cdots, \bm{G}_K]$ with $\bm{G}_k \in \mathbb{C}^{MN\times Q}$ referring to the channel between the $k^{\rm th}$ UE and the BS, for all $k = 1, \cdots, K$; the block diagonal matrix $\bm{V} \triangleq \text{diag}\left(\bm{v}_1,\cdots,\bm{v}_K\right)$ represents the precoder, where $\bm{v}_k\in \mathbb{C}^{Q\times1}$ is set as the right singular vector corresponding to the largest singular value of $\bm{G}_k$, and $\bm{H} \triangleq [\bm{h}_1, \bm{h}_2, \cdots, \bm{h}_{\Psi}]$ with $\bm{h}_\psi \in \mathbb C^{MN \times 1}$ being the channel between the $\psi^{\rm th}$ inter-cell interference and the BS, for all $\psi = 1, \cdots, \Psi$. Since the interferences from adjacent cells are much farther from the target BS than the signals transmitted by the intra-cell UEs, each interference is assumed to be radiated from a point source, independent of the number of Tx antennas. Moreover, $\bm{x} = [x_1, x_2, \cdots, x_K]^T$ and $\bm{s} = [s_1, s_2, \cdots, s_\Psi]^T$ in \eqref{Eq-1} denote the desired signals transmitted by $K$ UEs in the target cell and the unwanted interferences coming from $\Psi$ interferers in adjacent cells, respectively. We assume that $\mathbb{E}\{\bm{x}\bm{x}^H\} = P_{\rm U}\bm{I}_{K}$ and $\mathbb{E}\{\bm{s}\bm{s}^H\} = P_{\rm I}\bm{I}_{\Psi}$, where $P_{\rm U}$ and $P_{\rm I}$ denote the Tx power of intra-cell UEs and inter-cell interferences, respectively. Finally, the vector $\bm{z} \in \mathbb{C}^{MN\times 1}$ in \eqref{Eq-1} denotes a circularly symmetric additive white Gaussian noise (AWGN) with $\mathbb{E}\{\bm{z}\bm{z}^H\} = N_0\bm{I}_{MN}$, where $N_0$ indicates the power of noise.

The hybrid beamformer employed at the BS consists of an analog beamformer $\bm{F}_{\rm RF} \in \mathbb{C}^{MN \times N_{\rm RF}}$, cascaded with a digital one $\bm{F}_{\rm BB} \in \mathbb{C}^{N_{\rm RF} \times K}$. To sufficiently exploit the RF chain resources, it is assumed that $N_{\rm RF} = K$. In practice, $\bm{F}_{\rm BB}$ is implemented in baseband by a digital signal processor without any numerical constraint, whereas $\bm{F}_{\rm RF}$ is usually realized by a set of analog phase shifters, each of unit-modulus. Using this hybrid beamformer and by \eqref{Eq-1}, the post-beamforming signal at the BS, i.e., an estimate of the transmitted signal $\bm{x}$, is given by
\begin{equation} \label{Eq-2}
	\hat{\bm{x}} = \bm{F}_{\rm BB}^H\bm{F}_{\rm RF}^H\left(\bm{GVx}+\bm{Hs}+\bm{z}\right).
\end{equation}

\subsection{Millimetter Wave Channel Modeling} \label{Section_ChannelModel}	

\begin{figure}[!t]
	\centering
	\includegraphics [width=3.5in]{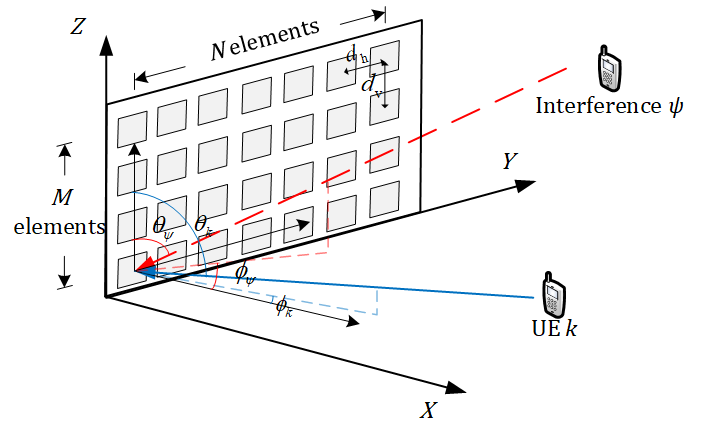}
	\caption{An illustrative UPA receiving the data from the desired user $k$, interfered with user $p$ in an adjacent cell.}
	\label{Fig_3Model}
\end{figure}

Figure~\ref{Fig_3Model} illustrates a UPA receiving the data from the desired user $k$, interfered with by user $p$ in an adjacent cell.  Due to the sparsity of mmWave propagation, the channel from the $k^{\rm th}$ UE, for all $k = 1, 2, \cdots, K$, to the BS can be conceived according to the extended Saleh-Valenzuela model with Rician fading \cite{38.901, 10214216}: 
\begin{equation}\label{Eq_channel-a}
	\bm{G}_k = \sum_{l=0}^{L_k-1}\alpha_{kl}\bm{a}_{\rm r}(\phi_{kl}^{\rm r},\theta_{kl}^{\rm r})\bm{a}_{\rm t}^H(\phi_{kl}^{\rm t}),
\end{equation}
where $L_k$ denotes the number of propagation paths from the $k^{\rm th}$ UE to the BS, and $\alpha_{kl}$ is the channel coefficient of the $l^{\rm th}$ path of the $k^{\rm th}$ UE. For the LoS path (i.e., $l=0$), $\alpha_{k0} = \sqrt{\kappa/(1+\kappa)} \exp\left({j\Phi_{k}^{\rm LoS}}\right)$, where $\kappa$ is the Rician $K$-factor and $\Phi_{k}^{\rm LoS}$ denotes a random initial phase. For the NLoS paths (i.e., $l = 1, \cdots, L_k-1$), the channel coefficient $\alpha_{kl} = \sqrt{1/((1+\kappa)(L_k-1)})\alpha'_{kl}$ with $\alpha'_{kl}\sim \mathcal{C} \mathcal{N}\left(0, 1\right)$. The vectors $\bm{a}_{\rm r}(\phi_{kl}^{\rm r}, \theta_{kl}^{\rm r})$ and $\bm{a}_{\rm t}(\phi_{kl}^{\rm t})$ in \eqref{Eq_channel-a} represent the Tx and Rx array steering vectors, respectively, with $\phi_{kl}^{\rm r}$, $\theta_{kl}^{\rm r}$, and $\phi_{kl}^{\rm t}$ denoting the horizontal AoA, vertical AoA, and angle of departure (AoD) of the $l^{\rm th}$ path of the $k^{\rm th}$ UE, respectively. 

As a UPA is employed at the BS, the receive array steering vector $\bm{a}_{\rm r}(\phi_{\rm r},\theta_{\rm r})$ with $\phi_{\rm r}$ and $\theta_{\rm r}$ being arbitrary horizontal and vertical AoAs, can be decomposed into the Kronecker product of horizontal steering vector $\bm{a}_{\rm h}(\Phi_{\rm r})$ and vertical steering vector $\bm{a}_{\rm v}(\Theta_{\rm r})$, that is, 
\begin{equation} \label{Eq_channel-b}
	\bm{a}_{\rm r}(\phi_{\rm r}, \theta_{\rm r}) = \bm{a}_{\rm h}(\Phi_{\rm r})\otimes\bm{a}_{\rm v}(\Theta_{\rm r}), 
\end{equation}
where
\begin{equation} \label{Eq_channel-c}
	\bm{a}_{\rm h}(\Phi_{\rm r}) = \left[1, e^{j\Phi_{\rm r}},\cdots,e^{j(N-1)\Phi_{\rm r}}\right]^T, 
\end{equation}
with $\Phi_{\rm r} \triangleq 2\pi d_{\rm h}\sin\theta_{\rm r}\sin\phi_{\rm r}/\lambda$, in which $d_{\rm h}$ denotes the horizontal spacing of antenna elements and $\lambda$ indicates the carrier wavelength, and
\begin{equation} \label{Eq_channel-d}
	\bm{a}_{\rm v}(\Theta_{\rm r}) = \left[1, e^{j\Theta_{\rm r}},\cdots,e^{j(M-1)\Theta_{\rm r}}\right]^T, 
\end{equation}
with $\Theta_{\rm r} \triangleq 2\pi d_{\rm v}\cos\theta_{\rm r}/\lambda$, in which $d_{\rm v}$ means the vertical spacing of antenna elements. Also, as a ULA is equipped at the UE, the Tx array steering vector is given by
\begin{equation}  \label{Eq_channel-e}
	\bm{a}_{\rm t}(\phi_{\rm t}) = \left[1, e^{j\frac{2\pi d_{\rm t}}{\lambda}\sin(\phi_{\rm t})},\cdots,e^{j\frac{2\pi(Q-1)d_{\rm t}}{\lambda}\sin(\phi_{\rm t})}\right]^T,
\end{equation}
where $d_{\rm t}$ is the spacing of antenna elements at each UE.
	
Like \eqref{Eq_channel-a}, the interference channel from the $\psi^{\rm th}$ interfering UE, for all $\psi = 1, 2, \cdots, \Psi$, to the BS is modeled in a multi-path way:
\begin{equation} \label{Eq_ICI}
	\bm{h}_{\psi} = \sum_{\gamma=1}^{\Gamma_{\psi}}\alpha_{\psi\gamma}\bm{a}_{\rm r}(\phi_{\psi\gamma}^{\rm r},\theta_{\psi\gamma}^{\rm r}),
\end{equation}
where $\alpha_{\psi\gamma} = \alpha'_{\psi\gamma}/\sqrt{\Gamma_{\psi}}$ denotes the channel coefficient with $\alpha'_{\psi\gamma} \sim \mathcal{C} \mathcal{N}\left(0, 1\right)$, and the steering vector $\bm{a}_{\rm r}(\phi_{\psi\gamma}^{\rm r},\theta_{\psi\gamma}^{\rm r})$ is defined similarly to \eqref{Eq_channel-b}, with horizontal AoA $\phi_{\psi\gamma}^{\rm r}$ and vertical AoA $\theta_{\psi\gamma}^{\rm r}$.

\section{Problem Formulation}
\label{Section_ProblemFormulation}
To maximize the achievable sum rate of the uplink transmission, in light of \eqref{Eq-2}, our task amounts to designing the analog beamformer $\bm{F}_{\rm RF}$ and digital beamformer $\bm{F}_{\rm BB}$. Unlike the pure digital beamformer, where both intra-cell and inter-cell interferences are dealt with jointly, the hybrid beamformer uses the analog beamformer to null inter-cell interferences. In contrast, the digital beamformer mitigates intra-cell interferences, thus fully exploiting the RF chain DoFs. The fundamental reason behind this function splitting is that one can readily obtain the AoAs rather than the full CSI of inter-cell interferences in real-world applications. As a consequence, the optimization problem for the maximal sum rate can be formulated as \cite{Zhu2017, 9217058}
\begin{subequations} \label{Eq-OPT}
	\begin{align}
		\mathcal{P}1:    &\	\max_{\bm{F}_{\rm BB}, \bm{F}_{\rm RF}} \; \sum_{k = 1}^{K}R_k \label{Eq_P1pro} \\
		\text{s.t.} 		& \ \left|\left[\bm{F}_{\rm RF}\right]_{i, j}\right|=1, \ \forall i,j \label{Eq_P1cons1} \\
		& \left\|\bm{F}_{\rm RF}^{H} \bm{h}_{\psi}\right\|^{2}=0, \ \forall \psi = 1, 2, \cdots, \Psi \label{Eq_P1cons2}
	\end{align}
\end{subequations}
where \eqref{Eq_P1cons1} reflects the unit-modulus constraints of the phase shifter network; \eqref{Eq_P1cons2} imposes the nulling of inter-cell interference completed by the analog beamformer, and $R_k$ in \eqref{Eq_P1pro} computes the Shannon capacity of the $k^{\rm th}$ as per
	\begin{equation}
		R_k = \log_2\left( 1+\frac{P_{\rm desired}^{(k)}}{P_{\rm intra}^{(k)}+P_{\rm inter}^{(k)}+ P_{\rm noise}^{(k)}} \right),
	\end{equation}
	where $P_{\rm desired}^{(k)} \triangleq P_{\rm U}|\bm{f}_{\rm BB}^H(k)\bm{F}_{\rm RF}^H\bm{G}_k\bm{v}_k|^2$, $P_{\rm intra}^{(k)} \triangleq \sum_{q\neq k}P_{\rm U}|\bm{f}_{\rm BB}^H(k)\bm{F}_{\rm RF}^H\bm{G}_q\bm{v}_q|^2$, $P_{\rm inter}^{(k)} \triangleq \sum_{\psi = 1}^{\Psi}P_{\rm I}|\bm{f}_{\rm BB}^H(k)\bm{F}_{\rm RF}^H\bm{h}_{\psi}|^2$, and $P_{\rm noise}^{(k)} \triangleq N_0\|\bm{f}_{\rm BB}^H(k)\bm{F}_{\rm RF}^H\|^2$, denoting the power of desired signal, intra-cell interference, inter-cell interference, and of noise, respectively, and $\bm{f}_{\rm BB}(k)$ refers to the $k^{\rm th}$ column of $\bm{F}_{\rm BB}$.

Due to the unit-modulus constraints outlined in \eqref{Eq_P1cons1}, finding a globally optimal solution to problem \(\mathcal{P}1\) is challenging from a mathematical perspective \cite{Ayach2014}. In our previous work \cite{Zhu2017}, we developed a hybrid beamformer using Kronecker decomposition for a linear antenna array and a single-path channel model of inter-cell interference. This optimization problem is similar to, but simpler than, the one described in \eqref{Eq-OPT}. The Kronecker hybrid beamformer demonstrates low computational complexity, effectively canceling inter-cell interferences while only requiring knowledge of their angles of arrival (AoAs).

However, the hybrid beamformer proposed in \cite{Zhu2017} cannot be directly applied to uniform planar arrays (UPAs) and does not account for the multi-path channel model for inter-cell interferences. Specifically, it fails to fully utilize the Kronecker factors, which significantly impairs the performance of the designed beamformer. Therefore, this paper focuses on UPAs with multi-path inter-cell interferences and exploits the primitive Kronecker decomposition. We propose a dynamic factor allocation method that markedly enhances the achievable sum rate.

\section{Hybrid Beamformer Design}
\label{Section_KroneckerBeamforming}
In this section, problem $\mathcal{P}1$ given by \eqref{Eq-OPT} is decoupled into two subproblems: the analog beamformer is designed to null inter-cell interferences and simultaneously to enhance desired signals, while the digital one is to mitigate intra-cell interferences. Although the decoupling of analog and digital beamformers makes the solution suboptimal, it makes the complicated problem tractable and the scheme implementable with low computational complexity and hardware cost. Simulation results to be discussed later show that the sum rate of the proposed scheme is close to that of the optimal digital MMSE scheme. 

To null inter-cell interferences and simultaneously enhance desired signals, the optimization problem for the analog beamformer $\bm{F}_{\rm RF}$ can be formulated as
\begin{subequations}  \label{Eq-P2}
	\begin{align} 
		\mathcal{P}2: & \ \max_{\bm{f}_{\rm RF}(k)}  \ |\bm{f}_{\rm RF}^H(k)\bm{G}_k\bm{v}_k|^2, \ k = 1, 2, \cdots, K,  \label{Eq-P2a} \\
		\text{s.t.}  & \ \left|\left[\bm{f}_{\rm RF}(k)\right]_{i}\right|=1, \ \forall i = 1, 2, \cdots, MN \label{Eq-P2b} \\
		& \ \left\|\bm{f}_{\rm RF}^{H}(k) \bm{h}_{\psi}\right\|^{2}=0, \ \forall \psi = 1, 2, \cdots, \Psi \label{Eq-P2c}
	\end{align}
\end{subequations}
where $\bm{f}_{\rm RF}(k)$ is the $k^{\rm th}$ column of $\bm{F}_{\rm RF}$. It is clear from \eqref{Eq-P2} that the analog beamforming vectors for different UEs are decoupled, i.e., $\bm{f}_{\rm RF}(k)$ is independent of $\bm{f}_{\rm RF}(j)$, $\forall k \ne j, k, j \in \{1, 2, \cdots, K\}$. Let $\bm{f}_{\rm RF}^{*}(k)$ be the optimal beamforming vector for the $k^{\rm th}$ UE, the analog beamformer can be shown as
\begin{equation} \label{Eq-12}
	\bm{F}_{\rm RF}^{*} = \left[\bm{f}_{\rm RF}^{*}(1), \bm{f}_{\rm RF}^{*}(2), \cdots, \bm{f}_{\rm RF}^{*}(K)\right].
\end{equation} 
Then, the digital beamformer is designed based on the equivalent channel $\tilde{\bm{G}} = (\bm{F}_{\rm RF}^{*})^H\bm{GV}$. In essence, the digital beamforming problem can be seen as the conventional unconstrained single-cell multi-user beamforming problem. Since the MMSE criterion has been proved optimal in the sense of information theory \cite[Sec. 8.3.4]{Tse2005}, the MMSE beamforming was widely applied to maximize the signal-to-interference-plus-noise ratio (SINR) of each UE \cite[Eq.(14)]{Zhu2017}. 

Next, the design of the analog beamformer (i.e., solving $\mathcal{P}2$ given by \eqref{Eq-P2}) is elaborated in three steps: primitive Kronecker decomposition, nulling inter-cell interferences by dynamic Kronecker factor allocation, and enhancing desired signals by using the remaining Kronecker factors, as each detailed below.
	
\subsection{Primitive Kronecker Decomposition of Steering Vectors}
\label{subSection_KroneckerDecomposition}
	Because the horizontal and vertical steering vectors given by \eqref{Eq_channel-c} and \eqref{Eq_channel-d}, respectively, have the Vandermonde structure and unit-modulus elements, we exploit the {\it primitive Kronecker decomposition} defined in Definition~\ref{Def-PKD} to maximize the number of Kronecker factors, that is, the value of $D$ in \eqref{Eq-10}. Concerning the UPA considered in this paper, we assume both $M$ and $N$ are even integers as a power of $2$, that is, $M \triangleq 2^{D_{\rm v}}$ and $N \triangleq 2^{D_{\rm h}}$, which are usually employed in real-world antenna configurations. Accordingly, the primitive decomposition of the horizontal and vertical steering vectors can be expressed as
	\begin{align}
		\lefteqn{\bm{a}_{\rm h}(\Phi_{\rm r})} \nonumber \\
		&= \left[1,\ e^{j\Phi_{\rm r} 2^{D_{\rm h}-1}}\right]^T \otimes \left[1,\ e^{j\Phi_{\rm r} 2^{D_{\rm h}-2}}\right]^T \otimes \cdots \otimes \left[1,\ e^{j\Phi_{\rm r}}\right]^T, \label{Eq_KPh}\\
		\lefteqn{\bm{a}_{\rm v}(\Theta_{\rm r})}  \nonumber \\
		&= \left[1,\ e^{j\Theta_{\rm r} 2^{D_{\rm v}-1}}\right]^T \otimes \left[1,\ e^{j\Theta_{\rm r}2^{D_{\rm v}-2}}\right]^T \otimes \cdots \otimes \left[1,\ e^{j\Theta_{\rm r}}\right]^T. \label{Eq_KPv}
	\end{align}
	Inserting \eqref{Eq_KPh}-\eqref{Eq_KPv} into \eqref{Eq_channel-b}, the UPA steering vector can be decomposed into
	\begin{align}
		\lefteqn{\bm{a}_{\rm r}(\phi_{\rm r},\theta_{\rm r})}  \nonumber \\
		&= \left[1,\ e^{j\Phi_{\rm r}2^{D_{\rm h}-1}}\right]^T  \otimes \cdots \otimes \left[1,\ e^{j\Phi_{\rm r}}\right]^T \otimes \left[1,\ e^{j\Theta_{\rm r}2^{D_{\rm v}-1}}\right]^T \nonumber \\
		&\quad \otimes \cdots \otimes \left[1,\ e^{j\Theta_{\rm r}}\right]^T \label{Eq_KPa0} \\
		&= \bm{a}^{(1)} \otimes \cdots \otimes \bm{a}^{(D_{\rm h})} \otimes \cdots \otimes \bm{a}^{(D)},  \label{Eq_KPa}
	\end{align}
	where $D \triangleq D_{\rm h}+D_{\rm v} = \log_2{MN}$. For ease of notation, we use $\bm{a}^{(d)}$ to denote the $d^{\rm th}$ factor in \eqref{Eq_KPa0}, for all $d = 1, 2, \cdots, D$. Then, substituting the corresponding AoAs into \eqref{Eq_KPa}, the steering vectors of the $\gamma^{\rm th}$ component from the $\psi^{\rm th}$ interference and $l^{\rm th}$ data channel regarding the $k^{\rm th}$ UE can be explicitly expressed as
	\begin{align}
		\bm{a}_{\rm r}(\phi^{\rm r}_{\psi\gamma},\theta^{\rm r}_{\psi\gamma}) &= \bm{a}^{(1)}_{\psi\gamma} \otimes  \cdots \otimes \bm{a}^{(D)}_{\psi\gamma}, \label{Eq_InterPhaseRes3} \\
		\bm{a}_{\rm r}(\phi^{\rm r}_{kl},\theta^{\rm r}_{kl}) &= \bm{a}^{(1)}_{kl} \otimes  \cdots \otimes \bm{a}^{(D)}_{kl}, \label{Eq_dataPhaseRes3}
	\end{align}
	where $\bm{a}^{(d)}_{\psi\gamma}$ and $\bm{a}^{(d)}_{kl}$ denote the $d^{\rm th}$ Kronecker factor of $\bm{a}_{\rm r}(\phi^{\rm r}_{\psi\gamma},\theta^{\rm r}_{\psi\gamma})$ and $\bm{a}_{\rm r}(\phi^{\rm r}_{kl},\theta^{\rm r}_{kl})$, respectively.
	
	Like \eqref{Eq_KPa}, let the analog beamformer $\bm{f}_{\mathrm {RF}}(k)$ to be designed have the same Kronecker decomposition:
	\begin{equation} \label{Eq_AnalogBeam3}
		\bm{f}_{\mathrm{RF}}(k) = \bm{f}^{(1)}(k) \otimes \bm{f}^{(2)}(k) \otimes \cdots \otimes \bm{f}^{(D)}(k),
	\end{equation}
	where $\bm{f}^{(d)}(k)$ denotes the $d^{\rm th}$ Kronecker factor of $\bm{f}_{\mathrm{RF}}(k)$, for all $d=1,\cdots , D$. By the definition of Kronecker product, if each element of $\bm{f}^{(d)}(k)$ is of unit-modulus, the constraint \eqref{Eq-P2b} is certainly satisfied.

\subsection{Nulling Inter-Cell Interference: Optimal Kronecker Factor Allocation}
\label{Subsection-KFA}
Now, we are in a position to resolve constraint \eqref{Eq-P2c}. Substituting \eqref{Eq_ICI} into \eqref{Eq-P2c} gives
\begin{equation} \label{Eq_interCanVec}
	\bm{f}_{\mathrm{RF}}^{H}(k) \bm{a}_{\rm r}(\phi^{\rm r}_{\psi\gamma},\theta^{\rm r}_{\psi\gamma}) = 0. 
\end{equation}
Then, inserting \eqref{Eq_InterPhaseRes3} and \eqref{Eq_AnalogBeam3} into \eqref{Eq_interCanVec} and recalling the mixed product rule of Kronecker product \cite[Eq. (2.11)]{Graham2018}, \eqref{Eq_interCanVec} can be equivalently rewritten as 
\begin{equation} \label{Eq_factorCanc}
	\left(\bm{f}^{(1)}(k)\right)^{H} \bm{a}_{\psi\gamma}^{(1)} \otimes \cdots \otimes\left(\bm{f}^{(D)}(k)\right)^{H} \bm{a}_{\psi\gamma}^{(D)} = 0,
\end{equation}
which implies the $\gamma^{\rm th}$ interference component from the $\psi^{\rm th}$ interfering UE can be fully canceled out by setting any factor in \eqref{Eq_factorCanc} to zero, that is,
\begin{equation} \label{Eq_interCan}
	\left(\bm{f}^{(d)}(k)\right)^{H} \bm{a}_{\psi\gamma}^{(d)} = 0, \, \forall d \in\{1,2, \cdots, D\}.
\end{equation}
It is noteworthy that the value of $D$ must be no smaller than $\Gamma+\lceil \log_{2}{K} \rceil$, that is, $D \geq \Gamma +\lceil\log_2 K\rceil$, where $\Gamma = \sum_{\psi=1}^{\Psi}\Gamma_{\psi}$ is the total number of inter-cell interference paths. This constraint is because, as implied by \eqref{Eq_factorCanc}, the analog beamformer needs at least $\Gamma$ factors to null the inter-cell interferences from $\Gamma$ distinct propagation paths. In contrast, the digital beamformer needs at least another $\lceil\log_2 K\rceil$ factors to separate $K$ intra-cell data streams (an in-depth analysis will shortly be performed in Section~\ref{Subsection_OptimalityAnalysis}). Obviously, \eqref{Eq_interCan} implies that $\bm{f}^{(d)}(k)$ is orthogonal to $\bm{a}_{\psi\gamma}^{(d)}$ and, together with the fact that each element of $\bm{a}_{\psi\gamma}^{(d)}$ is of unit-modulus, thus, the solution to \eqref{Eq_interCan} can be explicitly given by
\begin{equation} \label{Eq_designBeam}
	\bm{f}^{(d)}(k) = \operatorname{diag}\left(\bm{a}_{\psi\gamma}^{(d)}\right) \bm{t},
\end{equation}
where $\bm{t}^T = [1, -1]$. Clearly, each element of $\bm{f}^{(d)}(k)$ is of unit-modulus.  

As demonstrated in \eqref{Eq_interCanVec}-\eqref{Eq_factorCanc}, inter-cell interference can be canceled by any one of the $D$ Kronecker factors, provided that condition \eqref{Eq_interCan} is met. This leads to a natural question: if multiple factors satisfy \eqref{Eq_interCan} for a specific instance of inter-cell interference, which one should be selected? In our previous work \cite{Zhu2017}, we used these factors in a straightforward, sequential manner, which was simple and practical but also inefficient. 

Instead, our proposed approach is to select the factor that most enhances the desired signals. To achieve this, we employ a measure matrix that assesses the correlation between the Kronecker factors and the desired signals. Subsequently, we develop a dynamic factor allocation scheme to maximize the desired signal power based on this measure matrix, as detailed below.

Let the factor $\bm{f}_{\psi\gamma}^{(d)}$ be the $d^{\rm th}$ beamformer factor that is capable of nulling the $\gamma^{\rm th}$ interference component from the $\psi^{\rm th}$ interfering UE, that is, $(\bm{f}_{\psi\gamma}^{(d)})^H\bm{a}^{(d)}_{\psi\gamma} = 0$. We define the measure matrix $\bm{U}_k$ with the $(d, \varepsilon)^{\rm th}$ entry $u_{d\varepsilon}^{(k)}$, in which $n = \sum_{i=1}^{\psi-1}\Gamma_{i}+\gamma$, representing the cross-correlation between the data path factors of the $k^{\rm th}$ UE and the beamformer factor $\bm{f}_{\psi\gamma}^{(d)}$, that is,
\begin{equation} \label{Eq_index}
	u_{d\varepsilon}^{(k)} = \left|(\bm{f}_{\psi\gamma}^{(d)})^H\sum_{l=0}^{L_k-1}(\alpha_{kl}\bm{a}_{\rm t}^H(\phi_{kl}^{\rm t})\bm{v}_k)^{1/D}\bm{a}^{(d)}_{kl}\right|, 
\end{equation} 
where $d\in \mathcal{D}$ with $\mathcal{D}$ denoting the set of designed beamformer factors satisfying \eqref{Eq_interCan}, and $\varepsilon \in \mathcal{E}$ with $\mathcal{E}$ being the set of inter-cell interference components. Consequently, the dynamic Kronecker factor allocation problem is to find the first $\Gamma$ maximal $u_{d\varepsilon}^{(k)}$ with different $d$ about different interference component $\varepsilon$, for all $\varepsilon = 1, 2, \cdots, \Gamma$. Let the set $\mathcal{F}_k$ indicate the output of the Kronecker factor allocation scheme, which contains $\Gamma$ pairs of $\{ d,\varepsilon \}$ that means the $d^{\rm th}$ beamformer factor is allocated to null the $\varepsilon^{\rm th}$ interference component for the $k^{\rm th}$ UE. Then, the dynamic Kronecker factor allocation problem can be formulated as 
\begin{equation}
	\mathcal{P}3: \ \max_{\mathcal{F}_k} \sum_{\{d, \, \varepsilon\} \in \mathcal{F}_k} u_{d\varepsilon}^{(k)}.\\
\end{equation}

To solve problem $\mathcal{P}3$, an iterative approach is formalized in {\bf Algorithm~\ref{Algorithm-1}}. The main idea is that at each iteration, the $\{d, \varepsilon\}$ pair with the maximal $u_{d\varepsilon}^{(k)}$ is selected from the residual beamformer factor set $\mathcal{D}_{\rm Res}$ and interference component set $\mathcal{E}_{\rm Res}$ until all inter-cell interferences from $\Gamma$  propagation paths are canceled out.

\begin{algorithm}[t]
	\caption{Dynamic Kronecker Factor Allocation}
	\small
	\setstretch{1.0}
	\begin{algorithmic}[1]
		\REQUIRE The channel coefficient $\alpha_{kl}$, the array steering vector $\bm{a}_{\rm t}(\phi_{kl}^{\rm t})$, the precoder matrix $\bm{V}$, and all Kronecker factors of $\bm{f}_{\rm RF}(k)$, $\bm{a}_{\rm r}(\phi_{\psi\gamma}^{\rm r},\theta_{\psi\gamma}^{\rm r})$ and $\bm{a}_{\rm r}(\phi_{kl}^{\rm r},\theta_{kl}^{\rm r})$.
		\STATE Calculate the measure matrix $\bm{U}_k$ as per \eqref{Eq_index}. $\mathcal{D}_{\rm Res} = \mathcal{D}$, $\mathcal{E}_{\rm Res} = \mathcal{E}$, $\mathcal{F}_k = \emptyset$; 
		\WHILE{$\mathcal{E}_{\rm Res} \neq \emptyset$}
		\STATE $\{d', \varepsilon'\}= \arg \max\limits_{d,\varepsilon} u_{d\varepsilon}^{(k)}$, for all $ d \in \mathcal{D}_{\rm Res}$ and $\varepsilon \in \mathcal{E}_{\rm Res}$;
		\STATE $\mathcal{F}_k = \mathcal{F}_k \cup \{d', \varepsilon'\}$;
		\STATE $\bm{f}^{(d')}(k) = \bm{f}_{\psi'\gamma'}^{(d')}$, where $\varepsilon' = \sum_{i=1}^{\psi'-1}\Gamma_{i}+\gamma'$;
		\STATE $\mathcal{D}_{\rm Res} = \mathcal{D}_{\rm Res}\setminus\{d'\}$ and $\mathcal{E}_{\rm Res} = \mathcal{E}_{\rm Res}\setminus \{ \varepsilon' \}$;
		\ENDWHILE
		\ENSURE the $\Gamma$ beamformer factors $\bm{f}^{(d)}(k)$, $\forall \{d, \varepsilon\} \in \mathcal{F}_k$ .
	\end{algorithmic}
	\label{Algorithm-1}
\end{algorithm}

\subsection{Enhancing Desired Signal: Fully Exploring Remaining Kronecker Factors}
\label{Subsection_AnalogSignalEnhan}
Insofar, $\Gamma$ out of $D$ factors in \eqref{Eq_AnalogBeam3} are determined. Then, the remaining $D-\Gamma$ factors are designed to enhance the desired signals further. However, since the $D-\Gamma$ factors to be designed are not consecutive but randomly scattered, it is hard for their joint design, if possible. To proceed, we propose to divide the Kronecker decomposition of $\bm{f}_{\mathrm{RF}}(k)$ into two parts and rearrange the already designed $\Gamma$ factors in the first part, while the remaining $D-\Gamma$ factors are in the second part. This rearrangement can be iteratively performed as follows. By starting a pointer from the beginning and another pointer from the end of the Kronecker decomposition of $\bm{f}_{\rm RF}(k)$, if the former pinpoints an undesigned factor and the latter pinpoints a designed factor (i.e., one of the $\Gamma$ factors designed to null inter-cell interferences), we swap the two factors and calculate the permutation matrix according to the weak commutative law of Kronecker product, as stated in Lemma~\ref{lemma_infFactors} below. Otherwise, update the two pointers. The factors of $\bm{a}_{\rm r}(\phi^{\rm r}_{kl},\theta^{\rm r}_{kl})$ can be rearranged in a similar way. As a result, an algorithm for rearranging Kronecker factors is formalized in {\bf Algorithm~\ref{Algorithm-2}}.

\begin{lemma}[The weak commutative law of Kronecker product] 
	\label{lemma_infFactors}
	To swap any two factors $\bm{c}_1$ and $\bm{c}_2$ in a Kronecker product of $k \ge 3$ vectors, it holds that
	\begin{subequations}
		\begin{align} \label{Eq_corollary}
		& \lefteqn{\bm{s}_1 \otimes \cdots \otimes  \bm{s}_p \otimes \bm{c}_1 \otimes \bm{s}_{p+1}  \otimes \cdots \otimes \bm{s}_{q} \otimes \bm{c}_2 \otimes \bm{s}_{q+1} \otimes  \cdots \otimes \bm{s}_{k}} \nonumber \\
		&= \bm{P}\left(\bm{s}_1 \otimes \cdots \otimes  \bm{s}_p \otimes \bm{c}_2 \otimes \bm{s}_{p+1}  \otimes \cdots \otimes \bm{s}_{q} \otimes \bm{c}_1 \otimes \bm{s}_{q+1} \otimes \cdots \otimes \bm{s}_{k}\right), 
		\end{align}
		where the permutation matrix can be explicitly expressed as
		\begin{align} \label{Eq_corollary2}
			\bm{P} &=\sum_{i,j = 1}^{m_{c_1}, m_{c_2}} \bm{I}_{m_{s_1}} \otimes \cdots \otimes \bm{I}_{m_{s_p}} \otimes \bm{E}_{ij,m_{c_1}\times m_{c_2}} \otimes \bm{I}_{m_{s_{p+1}}} \otimes \cdots   \nonumber \\
			& \hspace{4em}  \otimes \bm{I}_{m_{s_{q}}} \otimes \bm{E}_{ji,m_{c_2}\times m_{c_1}}\otimes \bm{I}_{m_{s_{q+1}}}\otimes \cdots \otimes \bm{I}_{m_{s_k}}, 
		\end{align}
		with $m_{s} \triangleq {\rm length}(s)$.
	\end{subequations}
\end{lemma}

\begin{proof}
	This lemma can be proved similarly to Section 2.5 of \cite{Graham2018}, which is omitted due to limited space.
\end{proof}

	\begin{algorithm}[!t]
		\caption{The Rearrangement of Kronecker Factors}
		\small
		\begin{algorithmic}[1]
			\small
			\REQUIRE All Kronecker factors of $\bm{f}_{\rm RF}(k)$ and $\bm{a}_{\rm r}(\phi_{kl}^{\rm r},\theta_{kl}^{\rm r})$.
			\STATE Set pointer $i = 1$, $j = D$ and permutation matrix $\bm{P}_k=\bm{I}$; 
			\WHILE{$j>i$}
			\IF{$\bm{f}^{(i)}(k)$ has not been designed but $\bm{f}^{(j)}(k)$ has been designed}
			\STATE Exchange the sequence of $\bm{f}^{(i)}(k)$ and $\bm{f}^{(j)}(k)$;
			\STATE Exchange the sequence of $\bm{a}_{kl}^{(i)}$ and $\bm{a}_{kl}^{(j)}$ for all $l$;
			\STATE Calculate the permutation matrix $\bm{P}_{ij}$ pertaining to the exchange of the $i^{\rm th}$ and $j^{\rm th}$ Kronecker factors according to Lemma~\ref{lemma_infFactors};
			\STATE $\bm{P}_k \leftarrow \bm{P}_k\bm{P}_{ij}$;
			\ELSIF{$\bm{f}^{(i)}(k)$ and $\bm{f}^{(j)}(k)$ have been designed}
			\STATE $i \leftarrow i+1$;
			\ELSIF{$\bm{f}^{(i)}(k)$ and $\bm{f}^{(j)}(k)$ have not been designed}
			\STATE $j \leftarrow j-1$;
			\ELSE
			\STATE $i \leftarrow i+1$, $j \leftarrow j-1$;
			\ENDIF
			\ENDWHILE
			\ENSURE $\bm{P}_k$, $\bm{f}_{\rm RF}(k)$ and $\bm{a}(\phi_{kl}, \theta_{kl})$ with rearranged Kronecker factors.
		\end{algorithmic}
		\label{Algorithm-2}
	\end{algorithm}

Let $\bm{f}'_{\mathrm{RF}}(k)$ be the rearranged $\bm{f}_{\rm RF}(k)$, that is, 
\begin{align} \label{Eq-28}
	\bm{f}'_{\mathrm{RF}}(k) & \triangleq \underbrace{\bm{f}'^{(1)}(k) \otimes \bm{f}'^{(2)}(k) \otimes \cdots \otimes \bm{f}'^{(\Gamma)}(k)}_{\bm{f}'_{\Gamma}(k)} \otimes \nonumber \\
	&  \qquad \underbrace{\bm{f}'^{(\Gamma+1)}(k) \otimes \bm{f}'^{(\Gamma+2)}(k) \otimes \cdots \otimes \bm{f}'^{(D)}(k)}_{\bm{f}'_{\mathrm{Res}}(k)},
\end{align}
 then, the beamformer $\bm{f}_{\mathrm{RF}}(k)$ can be rewritten as
\begin{equation} \label{Eq-29}
	\bm{f}_{\mathrm{RF}}(k) = \bm{P}_k\bm{f}'_{\mathrm{RF}}(k) =\bm{P}_k\left\{\bm{f}'_{\Gamma}(k) \otimes \bm{f}'_{\mathrm{Res}}(k)\right\}, 
\end{equation}
where $\bm{P}_k$ can be readily determined by successively using Lemma \ref{lemma_infFactors}. Likewise, the steering vector $\bm{a}_{\rm r}(\phi^{\rm r}_{kl}, \theta^{\rm r}_{kl})$ can be decomposed into:
\begin{equation} \label{Eq-37}
	\bm{a}_{\rm r}(\phi^{\rm r}_{kl},\theta^{\rm r}_{kl}) = \bm{P}_k \left( \bm{a}'_{kl, \Gamma} \otimes \bm{a}'_{kl,\rm Res} \right), \quad l = 0, 1, \cdots, L_k-1,
\end{equation}
where $\bm{a}'_{kl, \Gamma}$ is the Kronecker product of the first $\Gamma$ rearranged factors and $\bm{a}'_{kl,\rm Res}$ is the Kronecker product of the remaining factors. Substituting \eqref{Eq-29} and \eqref{Eq-37} into \eqref{Eq-P2a} yields

\begin{align}
\lefteqn{ \left|\bm{f}_{\rm RF}^H(k)\bm{G}_k\bm{v}_k\right|^2 } \nonumber \\
	&= \left|\sum_{l=0}^{L_k-1}\tilde{\alpha}_{kl}\left( \bm{f}'_{\Gamma}(k) \otimes \bm{f}'_{\rm Res}(k)\right)^H\bm{P}_k^H\bm{P}_k \left( \bm{a}'_{kl,\Gamma} \otimes \bm{a}'_{kl,\rm Res} \right) \right|^2 \label{Eq_fgTrans1} \\
	&=  \Bigg|(\bm{f}'_{\rm Res}(k))^H\underbrace{\sum_{l=0}^{L_k-1}\tilde{\alpha}_{kl}\left( (\bm{f}'_{\Gamma}(k))^H \bm{a}'_{kl,\Gamma}\right)\bm{a}'_{kl,\rm Res}}_{\tilde{\bm{g}}_k} \Bigg|^2, \label{Eq_fgTrans2}
\end{align} 
where $\tilde{\alpha}_{kl} = \alpha_{kl}\bm{a}_{\rm t}^H(\phi_{kl}^{\rm t})\bm{v}_k $. The equality $\bm{P}_k^H\bm{P}_k = \bm{I}$ and the mixed product rule of Kronecker product are exploited to derive \eqref{Eq_fgTrans2}.

With the already determined $\Gamma$ factors of $\bm{f}_{\mathrm{RF}}(k)$ and in view of \eqref{Eq_fgTrans1}-\eqref{Eq_fgTrans2}, the problem $\mathcal{P}2$ can be simplified as
\begin{align}
	\mathcal{P}4: & \ \max_{\bm{f}'_{\rm Res}(k)} \ {\left|(\bm{f}'_{\rm Res}(k))^H \tilde{\bm{g}}_k\right|^{2} }\\
	  {\text { s.t. }} & \ {\left| \left[ \bm{f}'_{\rm Res}(k) \right]_{i} \right|=1, \ \forall i.}
\end{align}
Clearly, the solution to $\mathcal{P}4$ is \cite{9217058}
\begin{equation} \label{Eq_signalEhance}
	\bm{f}'_{\rm Res}(k) = \exp\left(j \, {\rm angle}(\tilde{\bm{g}}_k)\right).
\end{equation}
Finally, substituting \eqref{Eq_signalEhance} into \eqref{Eq-29} gives the desired $\bm{f}_{\mathrm{RF}}(k)$. To sum up, the proposed hybrid beamformer based on primitive Kronecker decomposition is formalized in {\bf Algorithm~\ref{Algorithm-3}}. 

\begin{algorithm}[t]
	\caption{The Hybrid Beamformer Based on the Primitive Kronecker Decomposition}
	\small
	\begin{algorithmic}[1]
		\REQUIRE The data channel AoAs $\phi_{kl}^{\rm r}$ and $\theta_{kl}^{\rm r}$, AoDs $\phi_{kl}^{\rm t}$, and channel gains $\alpha_{kl}$, for all $k= 1, \cdots, K$ and $ l = 0, \cdots, L_k-1$, the precoder $\bm{V}$, and inter-cell interference channel AoAs $\phi_{\psi\gamma}^{\rm r}$ and $\theta_{\psi\gamma}^{\rm r}$, for all $\psi = 1, \cdots, \Psi$ and $\gamma = 1, \cdots, \Gamma_{\psi}$.
		\FOR {$k = 1,\cdots,K$}
			\STATE Decompose the steering vectors of interference and data path of the $k^{\rm th}$ UE according to \eqref{Eq_InterPhaseRes3} and \eqref{Eq_dataPhaseRes3}, respectively, and express $\bm{f}_{\rm RF}(k)$ as \eqref{Eq_AnalogBeam3}; 
			\STATE Determine the $\Gamma$ Kronecker factors of $\bm{f}_{\rm RF}(k)$ to null inter-cell interferences by using Algorithm~\ref{Algorithm-1};
			\STATE Rearrange the Kronecker factors of $\bm{f}_{\rm RF}(k)$ and $\bm{a}_{\rm r}(\phi_{kl}^{\rm r}, \theta_{kl}^{\rm r})$ according to Algorithm~\ref{Algorithm-2};
			\STATE Determine $\bm{f}'_{\rm Res}(k)$ to enhance the desired signals as per \eqref{Eq_signalEhance};
			\STATE Calculate $\bm{f}_{\rm RF}(k)$ according to \eqref{Eq-29};
		\ENDFOR
		\STATE Calculate $\bm{F}_{\rm BB}$.
		\ENSURE The analog beamformer $\bm{F}_{\rm RF}$ and digital beamformer $\bm{F}_{\rm BB}$.
	\end{algorithmic}
	\label{Algorithm-3}
\end{algorithm}

In practice, as the analog beamformer is usually implemented by a set of phase shifters whose phase-shifting accuracy is much lower than that of the digital beamformer, it is unnecessary to adjust the analog and digital beamformers simultaneously. In theory, the change of the AoAs of inter-cell interferences is usually much slower than that of the CSI of intra-cell users due to their relatively far distances to the target BS. By jointly accounting for these two facts, a low-complexity hybrid beamformer is developed below, in which only the AoAs of both the data and interference channels are used to design the analog beamformer. In contrast, the full CSI of data channels determines the digital one.

\subsection{A Low-Complexity Hybrid Beamformer}
The low-complexity hybrid beamformer adjusts the analog and digital beamformers at different speeds: only if the AoA of any data or interference channels changes, the analog beamformer is recalculated by the proposed analog beamformer in this subsection; otherwise, we need only to calculate the digital beamformer. To be specific, compared with Algorithm~\ref{Algorithm-3}, the low-complexity hybrid beamformer has three major differences: 
\begin{enumerate}
	\item [1)] Unlike \eqref{Eq_dataPhaseRes3}, where all received array steering vectors of the data channels are decomposed, only the array steering vector about the LoS path (i.e., $l = 0$) from the $k^{\rm th}$ UE is decomposed as
	\begin{equation}
		\bm{a}_{\rm r}(\phi^{\rm r}_{k0},\theta^{\rm r}_{k0}) = \bm{a}^{(1)}_{k0} \otimes  \cdots \otimes \bm{a}^{(D)}_{k0}; \label{Eq_dataPhaseRes4}
	\end{equation}
	\item [2)] The measure matrix $\bm{U}_k$ represents the cross-correlation between the beamformer factor $\bm{f}_{\psi\gamma}^{(d)}$ and the factor of LoS path (rather than all paths in \eqref{Eq_index}) for the $k^{\rm th}$ UE, that is,
	\begin{equation} \label{Eq_indexdiffscale}
		u_{d\varepsilon}^{(k)} = \left|(\bm{f}_{\psi\gamma}^{(d)})^H\bm{a}^{(d)}_{k0}\right|;
	\end{equation} 
	\item [3)] The remaining factors enhance the LoS signal. Accordingly, \eqref{Eq_fgTrans1} is recomputed as
	\begin{align} 
		\lefteqn{ \left|\bm{f}_{\rm RF}^H(k)\bm{g}_k\right|^2} \nonumber \\
		&\approx \left|\bm{f}_{\rm RF}^H(k)(\alpha_{k0}\bm{a}_{\rm t}^H(\phi_{k0}^{\rm t})\bm{v}_k)\bm{a}_{\rm r}(\phi_{k0}^{\rm r},\theta_{k0}^{\rm r})\right|^2 \nonumber \\
		&= \Big|(\alpha_{k0}\bm{a}_{\rm t}^H(\phi_{k0}^{\rm t})\bm{v}_k)\left( \bm{f}'_{\Gamma}(k) \otimes \bm{f}'_{\rm Res}(k)\right)^H \nonumber \\
		&\quad \times \bm{P}_k^H\bm{P}_k \left( \bm{a}'_{k0,\Gamma} \otimes \bm{a}'_{k0,\rm Res} \right) \Big|^2 \nonumber \\
		&=  \Big|(\alpha_{k0}\bm{a}_{\rm t}^H(\phi_{k0}^{\rm t})\bm{v}_k)(\bm{f}'_{\rm Res}(k))^H \nonumber \\
		&\quad \times \underbrace{\left( (\bm{f}'_{\Gamma}(k))^H \bm{a}'_{k0,\Gamma}\right)\bm{a}'_{k0,\rm Res}}_{\tilde{\bm{g}}_k'} \Big|^2. \label{Eq_fgTransDiffScale3}
	\end{align} 
Like \eqref{Eq_signalEhance}, the remaining factors is determined by
	\begin{equation} \label{Eq_signalEhanceDiff}
		\bm{f}'_{\rm Res}(k) = \exp\left({j \, \rm angle}(\tilde{\bm{g}}_k')\right).
	\end{equation}
\end{enumerate}
In sum, the proposed low-complexity hybrid beamformer is formalized in {\bf Algorithm \ref{Algorithm-4}}.

	\begin{algorithm}[t]
		\small
		\caption{A Low-Complexity Hybrid Beamformer}
		\begin{algorithmic}[1]
			\REQUIRE The data channel AoAs $\phi_{k0}$ and $\theta_{k0}$, for all $k= 1, \cdots, K$, and inter-cell interference channel AoAs $\phi_{\psi\gamma}$ and $\theta_{\psi\gamma}$, for all $\psi = 1, \cdots, \Psi$ and $\gamma = 1, \cdots, \Gamma_{\psi}$.
			\IF{The AoA of data or interference channel is changed}
			\FOR {$k = 1,\cdots,K$}
			\STATE Decompose the steering vectors of interference and data path of the $k^{\rm th}$ UE according to \eqref{Eq_InterPhaseRes3} and \eqref{Eq_dataPhaseRes4}, respectively, and express $\bm{f}_{\rm RF}(k)$ as \eqref{Eq_AnalogBeam3}; 
			\STATE Determine the $\Gamma$ Kronecker factors of $\bm{f}_{\rm RF}(k)$ to null inter-cell interferences by using Algorithm~\ref{Algorithm-1} with measure matrix \eqref{Eq_indexdiffscale};
			\STATE Rearrange the Kronecker factors of $\bm{f}_{\rm RF}(k)$ and $\bm{a}(\phi_{k0}, \theta_{k0})$ according to Algorithm~\ref{Algorithm-2};
			\STATE Determine $\bm{f}'_{\rm Res}(k)$ to enhance the desired signals as per \eqref{Eq_signalEhanceDiff};
			\STATE Calculate $\bm{f}_{\rm RF}(k)$ according to \eqref{Eq-29};
			\ENDFOR
			\ENDIF
			\STATE Calculate $\bm{F}_{\rm BB}$.
			\ENSURE The analog beamformer $\bm{F}_{\rm RF}$ and digital beamformer $\bm{F}_{\rm BB}$.
		\end{algorithmic}
		\label{Algorithm-4}
\end{algorithm}

\section{Optimality and Complexity Analyses}
\label{Section_OptimalityComputationalComplexity}
This section performs an optimality analysis from a subspace perspective to demonstrate the excellent efficiency of the proposed dynamic Kronecker factor allocation. It is followed by sufficient antenna configuration conditions to implement the proposed beamformers and complexity analysis.

\subsection{Optimality Analysis of Dynamic Kronecker Factor Allocation}
\label{Subsection_OptimalityAnalysis}

Unlike the pure digital MMSE beamformer that finds an optimal tradeoff between mitigating interference-plus-noise and enhancing the desired signal \cite[Sec. 4.1]{Bjornson2017}, the two proposed schemes aim first to eliminate inter-cell interferences and then maximize the instantaneous signal-to-noise ratio (SNR). Moreover, the primitive Kronecker decomposition is applied to more efficiently exploit the spatial DoFs of massive UPA so that the same interference can be fully eliminated in a lower-dimensional space. Specifically, as shown in \eqref{Eq-P2b}-\eqref{Eq-P2c}, the analog beamforming vector $\bm{f}_{\rm RF}(k) \in \mathbb{C}^{MN}$ aiming to eliminate the $\psi^{\rm th}$ interference is originally on the complex circle manifold of $\mathbb{C}^{MN}$ \cite{Yu2016, Alhujaili2019}. By using primitive Kronecker decomposition, the constraint \eqref{Eq-P2c} is equivalently rewritten as \eqref{Eq_factorCanc}, where the $d^{\rm th}$ Kronecker factor $\bm{f}^{(d)}(k) \in \mathbb{C}^{n_d}$ is capable of nulling the $\gamma^{\rm th}$ component of the $\psi^{\rm th}$ interference. As $n_d \ll MN$, for all $d = 1,\cdots, D$, primitive Kronecker decomposition enables interference elimination in a much lower-dimensional space. However, since there is more than one Kronecker factor that is capable of eliminating the $\gamma^{\rm th}$ component of the ${\psi}^{\rm th}$ interference, which one should we prefer?

\begin{figure*}[!t] 
	\centering 
	\subfigure[The 1st factor $\bm{f}_{\psi\gamma}^{(1)}$]{\includegraphics[width = 0.3\textwidth]{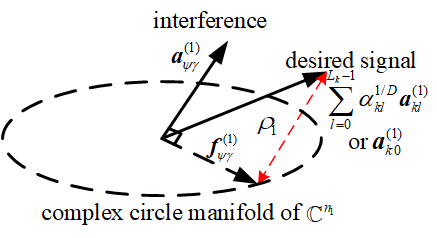} \label{Fig_2a}}
	\subfigure[The 2nd factor $\bm{f}_{\psi\gamma}^{(2)}$]{\includegraphics[width = 0.3\textwidth]{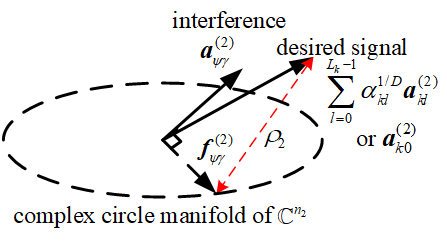} \label{Fig_2b}} 
	\subfigure[The 3rd factor $\bm{f}_{\psi\gamma}^{(3)}$]{\includegraphics[width = 0.3\textwidth]{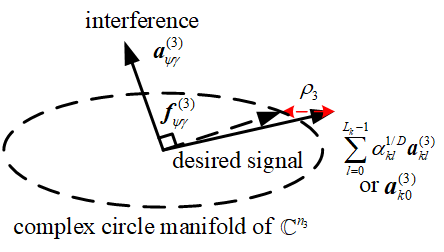} \label{Fig_2c}} 
	\caption{An illustrative example of how to determine the optimal Kronecker factor for a given beamforming vector $\bm{f}_{\rm RF}(k)$. Although all three factors can effectively null the $i^{\rm th}$ inter-cell interference because the $i^{\rm th}$ factor $\bm{f}_{\psi\gamma}^{(i)}$ of the beamforming vector is orthogonal to the $i^{\rm th}$ factor $\bm{a}_{\psi\gamma}^{(i)}$ of the interference signal, $i = 1, 2, 3$, the 3rd factor $\bm{f}_{\psi\gamma}^{(3)}$ is preferred to $\bm{f}_{\psi\gamma}^{(1)}$ and $\bm{f}_{\psi\gamma}^{(2)}$, due to its capability of enhancing the desired signal (the distances $\rho_3 < \rho_1 < \rho_2$).}
	\label{OptimizationExplanation}		
\end{figure*}

For illustration purposes, Fig.~\ref{OptimizationExplanation} shows three Kronecker factors $\bm{f}_{\psi\gamma}^{(1)}$, $\bm{f}_{\psi\gamma}^{(2)}$ and $\bm{f}_{\psi\gamma}^{(3)}$ of the beamforming vector $\bm{f}_{\mathrm {RF}}(k)$ that are orthogonal to three interference factors $\bm{a}_{\psi\gamma}^{(1)}$, $\bm{a}_{\psi\gamma}^{(2)}$ and $\bm{a}_{\psi\gamma}^{(3)}$ of the $\gamma^{\rm th}$ component of the $\psi^{\rm th}$ interference, respectively. According to \eqref{Eq_factorCanc}, each pair can eliminate the $\gamma^{\rm th}$ component; thus, these three Kronecker factors perform equally if only interference elimination is concerned. This fact was applied in \cite{Zhu2017}, where the Kronecker factors of $\bm{f}_{\mathrm {RF}}(k)$ are used in natural sequence to null inter-cell interference. This method is straightforward and effective, yet inefficient. In our design, however, we further exploit the feature that these factors have different distances from the resulting desired signal, as shown in Fig.~\ref{OptimizationExplanation} where the distances $\rho_3 < \rho_1 < \rho_2$. As a result, in addition to interference mitigation, the capability of signal enhancement is accounted for in our design and, therefore, the 3rd Kronecker factor $\bm{f}_{\psi\gamma}^{(3)}$ is preferred to $\bm{f}_{\psi\gamma}^{(1)}$ and $\bm{f}_{\psi\gamma}^{(2)}$ due to $\rho_3 < \rho_1 < \rho_2$. This is exactly the rationale behind Algorithm~\ref{Algorithm-1}. After the dynamic factor allocation as per Algorithm~\ref{Algorithm-1}, it is evident that the $\Gamma$ Kronecker factors to eliminate all inter-cell interferences from $\Gamma$  propagation paths may not be consecutive but randomly scattered. Consequently, Algorithm~\ref{Algorithm-2} is developed to rearrange these factors so that the resulting beamforming vector can be divided into two consecutive parts: one for interference elimination and the other for subsequent signal enhancement, as shown in the derivations from \eqref{Eq-28} to \eqref{Eq_fgTrans2}.

\subsection{Optimal Antenna Configuration: A Sufficient Condition}
The following theorem provides sufficient conditions for the optimal antenna configuration to enable dynamic factor allocation in practice.

\begin{theorem} \label{theorem_antennaNumber}
	Given a massive UPA of size $M \times N$, let the prime factorization of the product of $M$ and $N$ be $MN = n_1 n_2  \cdots n_D$, with $n_1 \leq n_2 \leq \cdots  \leq n_D$. If the partial product $n_1 n_2 \cdots n_{D-\Gamma} \geq K$, then the two proposed algorithms can not only eliminate the inter-cell interferences from $\Gamma$  propagation paths but also enhance desired signals of intra-cell UEs, as well as perform the MMSE detection of $K$ intra-cell UEs. Furthermore, to fully exploit the spatial DoFs, the optimal antenna configuration is $MN = 2^{\Gamma + \lceil \log_{2}{K} \rceil}$, that is, $D = \Gamma + \lceil \log_{2}{K} \rceil$.
\end{theorem}

\begin{proof}
	By \eqref{Eq-12}, the MMSE digital beamformer is designed as per the equivalent channel $\tilde{\bm{G}} = (\bm{F}_{\rm RF}^{*})^H\bm{GV}$. To separate the signals of $K$ intra-cell UEs, it is necessary that ${\rm rank}(\tilde{\bm{G}}) = K$. Since ${\rm rank}(\bm{GV}) = K$ holds in practice, the requirement ${\rm rank}(\tilde{\bm{G}}) = K$ reduces to ${\rm rank}(\bm{F}_{\rm RF}^{*}) = K$, by recalling the fact that ${\rm rank}(\bm{AB}) = {\rm rank}(\bm{A})$ if $\bm{B}$ is right invertible \cite[Prop. 2.6.3.]{Bernstein2009}. Thus, the analog beamforming vectors $\bm{f}_{\rm RF}(k)$, $k=1,\cdots,K,$ must be linearly independent.
	
	Consider the extreme case that all $K$ analog beamforming vectors choose the same Kronecker factor to cancel inter-cell interference, i.e., $\bm{P}_1 = \bm{P}_2 = \cdots = \bm{P}_K \triangleq \bm{P}$ and $\bm{f}'_{\Gamma}(1) = \bm{f}'_{\Gamma}(2) = \cdots = \bm{f}'_{\Gamma}(K) \triangleq \bm{f}'_{\Gamma}$, then, \eqref{Eq-29} can be rewritten as 
	\begin{equation} \label{Eq-45}
		\bm{f}_{\mathrm{RF}}(k) = \bm{P}\left\{\bm{f}'_{\Gamma} \otimes \bm{f}'_{\mathrm{Res}}(k)\right\}.
	\end{equation}
	Then, the linear independence of analog beamforming vectors implies that
	\begin{equation} \label{Eq-46}
		\sum_{k=1}^{K}\omega_k\bm{f}_{\mathrm{RF}}(k)=\bm{0}, \text{ if and only  if } \bm{\omega} = \bm{0},
	\end{equation}
	where $\bm{\omega} = [\omega_1, \omega_2, \cdots, \omega_K]^T$. Inserting \eqref{Eq-45} into \eqref{Eq-46} yields 
	\begin{equation}
		\bm{P}\left\{\bm{f}'_{\Gamma} \otimes \sum_{k=1}^{K}\omega_k\bm{f}'_{\mathrm{Res}}(k)\right\}=\bm{0}, \text{ if and only  if } \bm{\omega} = \bm{0},
	\end{equation}
	which means that $\bm{f}'_{\mathrm{Res}}(1),\bm{f}'_{\mathrm{Res}}(2), \cdots, \bm{f}'_{\mathrm{Res}}(K)$ are linearly independent. This suggests that the length of $\bm{f}'_{\mathrm{Res}}(k)$ must be greater than or equal to $K$, that is,
	\begin{equation} \label{Eq-47}
		{\rm length}(\bm{f}'_{\mathrm{Res}}(k)) \geq K, \ \forall k = 1, \cdots, K.
	\end{equation} 
	In practice, the exact value of ${\rm length}(\bm{f}'_{\mathrm{Res}}(k))$ depends on which factors are used to eliminate inter-cell interferences. Two extreme cases are as follows:
	\begin{itemize}
		\item[1)] If the first $\Gamma$ Kronecker factors of $\bm{f}^{(d)}(k)$ are used for interference elimination, the remaining spatial DoFs for signal enhancement is $n_{\Gamma+1}n_{\Gamma+2} \cdots n_{D}$;	
		\item[2)] If the last $\Gamma$ Kronecker factors of $\bm{f}^{(d)}(k)$ are used for interference elimination, the remaining spatial DoFs for signal enhancement is $n_{1}n_{2} \cdots n_{D-\Gamma}$.
	\end{itemize}
	Combining the above two extreme cases gives
	\begin{equation} \label{Eq-48}
		n_1 \cdots n_{D-\Gamma} \leq {\rm length}(\bm{f}'_{\mathrm{Res}}(k)) \leq n_{\Gamma+1}\cdots n_{D}.
	\end{equation}
	By \eqref{Eq-47} and \eqref{Eq-48}, we infer that if $n_1n_2 \cdots n_{D-\Gamma} \ge K$, the proposed algorithms can be efficiently applied for both interference elimination and signal enhancement. 
	
	Finally, by recalling the principle of prime factorization, a primitive Kronecker decomposition implies $n_1 = n_2 = \cdots = n_D = 2$, that is, $MN = 2^D$. In this case, there is the maximum number of Kronecker factors. Among them, whichever factors are selected for interference elimination, we have ${\rm length}(\bm{f}'_{\Gamma}) = 2^\Gamma$, which is indeed the minor spatial DoFs used for eliminating the inter-cell interferences from $\Gamma$  propagation paths. Then, the remaining DoFs are used for signal enhancement, requiring $2^{D-\Gamma} \geq K$ as discussed above. Therefore, we deduce that the optimal antenna configuration is $MN = 2^{\Gamma+\lceil \log_{2}{K} \rceil}$. This completes the proof.	
\end{proof}

\subsection{Complexity Analysis}
Now, we evaluate the computational complexity of the two proposed algorithms. In particular, the complexity of computing an analog beamformer consists of two parts: One is to construct the measure matrix according to \eqref{Eq_index}, which yields the complexity $\mathcal{O}\left(K\Gamma\log_2(MN)\right)$. The other includes the signal enhancement given by \eqref{Eq_signalEhance} and the combination of Kronecker factors as per \eqref{Eq-37}, introducing $\mathcal{O}\left(KMN\right)$ complexity. Calculating the permutation matrix in  Algorithm~\ref{Algorithm-2} can be achieved by simple logical operations, such that the complexity is negligible. The optimal MMSE beamformer's complexity is widely known to be $\mathcal{O}\left(K^3\right)$. To sum up, the overall computational complexity of Algorithm~\ref{Algorithm-3} is $\mathcal{O}\left(K\Gamma\log_2(MN)+KMN+K^3\right)$. 

The computational complexity of Algorithm~\ref{Algorithm-4} can be analyzed similarly to previous algorithms. It is given by $\mathcal{O}\left(\eta (K\Gamma\log_2(MN)+KMN) + K^3\right)$, where $\eta \le 1$ is the ratio of the adjustment speed of the analog beamformer to that of the digital one. In the context of massive MIMO, where $M$ and $N$ are large, and considering that $\Gamma \ll M$ and $\Gamma \ll N$, the complexities of Algorithms~\ref{Algorithm-3} and \ref{Algorithm-4} be approximated as follows: \(\mathcal{O}\left(KMN + K^3\right)\) for Algorithm \ref{Algorithm-3} and \(\mathcal{O}\left(\eta KMN + K^3\right)\) for Algorithm \ref{Algorithm-4}. In scenarios involving low mobility, where $\eta \ll 1$, the complexity of Algorithm \ref{Algorithm-4} simplifies to $\mathcal{O}\left(K^3\right)$, independent of antenna size.

For comparison purposes, the computational complexity of the successive KHB scheme developed in our previous work \cite{Zhu2017} and the pure digital MMSE scheme are precisely $\mathcal{O}\left(KMN+K^3\right)$ and $\mathcal{O}\left((MN)^3\right)$, respectively. Although the pure digital MMSE scheme is the optimal unconstrained linear beamforming algorithm \cite[Chap. 6]{Verdu1998}, its computational complexity and hardware cost are much higher than those of the other three Kronecker hybrid beamformers.In conclusion, Table~\ref{table_Complexity} compares the computational complexity of the proposed schemes with that of the benchmark schemes. It is observed that while the proposed Algorithm~\ref{Algorithm-3} has higher computational complexity than the benchmark scheme in [20], Algorithm~\ref{Algorithm-4} has lower complexity. All proposed algorithms exhibit lower complexity than the optimal MMSE scheme.

\begin{table}[!t]
	\centering  
	\caption{Computational Complexity Analysis}	
	\small			
	\begin{tabular}{cc}
		\toprule[1.3pt]
		{\bf Beamformer} &  {\bf Computational Complexity} \\
		\hline
		Algorithm \ref{Algorithm-3}        	   	& $\mathcal{O}\left(K\Gamma\log_2(MN)+KMN+K^3\right)$  \\
		Algorithm \ref{Algorithm-4}        	   	& $\mathcal{O}\left(K^3\right)$  \\
		Successive KHB \cite{Zhu2017}   	& $\mathcal{O}\left(KMN+K^3\right)$ 	    \\
		Optimal MMSE  	& $\mathcal{O}\left((MN)^3\right)$ 	 \\
		\bottomrule[1.3pt]  
		\label{table_Complexity}
	\end{tabular}	
\end{table}

\section{Simulation Results and Discussions}
\label{Section_SimulationResults}

This section presents and discusses the Monte-Carlo simulation results of the proposed beamformer compared to several typical beamformers reported in the literature. In line with the latest 3GPP TR 38.901 \cite{38.901}, the main simulation parameters are listed in Table \ref{Table_Parameter}, and each simulation is conducted with $5 \times 10^4$ Monte-Carlo runs. UE position determines the AoA and AoD of the LoS path of a UE channel. Inter-cell interferences' horizontal and vertical AoAs are assumed to follow the uniform distribution within $[0, 2\pi)$. Perfect CSI of intra-cell UEs and AoAs of inter-cell interferences are supposed to be available at the BS through a genius estimation algorithm (see, e.g., \cite{Fan2018}). Finally, the channel matrices are generated according to the mmWave channel model described in Section~\ref{Section_ChannelModel}. 

\begin{table}[!t]
\small
	\centering  
	\caption{Simulation Parameter Setting}				
	\begin{tabular}{cc}
		\toprule[1.3pt]
		{\bf Parameter} &  {\bf Value} \\
		\hline 
		Cell layout & Hexagonal grid \\
		Radius of cell & $100$ m \\
		Rician $K$-factor & $5$ dB\\
		Horizontal angular spread of NLoS path & $\pi$ \\
		Vertical angular spread of NLoS path & $\pi/2$ \\
		BS height & $10$ m \\
		UE height & $1.5-22.5$ m \\
		Number of UEs ($K$) & $4$ \\
		Number of UE antennas ($Q$) & $2$\\
		Antenna spacing ($d_{\rm t}$, $d_{\rm h}$, $d_{\rm v}$) & $\lambda/2$\\
		Number of propagation paths of each UE ($L_k$) & $2$ \\
		\bottomrule[1.3pt]  
		\label{Table_Parameter}
	\end{tabular}		
\end{table} 

Apart from the proposed algorithms, for comparison purposes, five other typical beamforming algorithms are evaluated in our simulation experiments, including the optimal pure digital MMSE beamforming, the exhaustive search scheme, the successive KHB \cite{Zhu2017}, the Broyden-Fletcher-Goldfarb-Shanno (BFGS)-based algorithm \cite{Jin2018}, and the equal gain combining-MMSE (EGC-MMSE) beamforming. Their main features are as follows:
\begin{enumerate}
	\item[1)] The MMSE scheme is an optimal performance benchmark. It features a fully digital architecture and achieves optimality due to being information lossless \cite[Sec. 8.3.4]{Tse2005}, although it entails the highest computational complexity and hardware costs;
	\item[2)] The exhaustive search scheme finds the best Kronecker factors to cancel the inter-cell interferences and maximize the desired signals by exhaustive searching, which serves as a benchmark to evaluate the dynamic Kronecker factor allocation scheme;
	\item[3)] The successive KHB scheme in \cite{Zhu2017} combines Kronecker analog beamforming and small-scale MMSE digital beamforming without accounting for Kronecker factor allocation;
	\item[4)] The BFGS algorithm in \cite{Jin2018} aims to minimize the Euclidean distance between the hybrid beamformer and the optimal MMSE beamformer;
	\item[5)] The EGC-MMSE scheme comprises an EGC analog beamformer and a small-scale MMSE digital one. 
\end{enumerate}
Among these methods, the optimal MMSE scheme uses a fully digital architecture, whereas the others use a hybrid one. 

\begin{figure}[!t]
	\centering
	\includegraphics [width = 0.475\textwidth,clip,keepaspectratio]{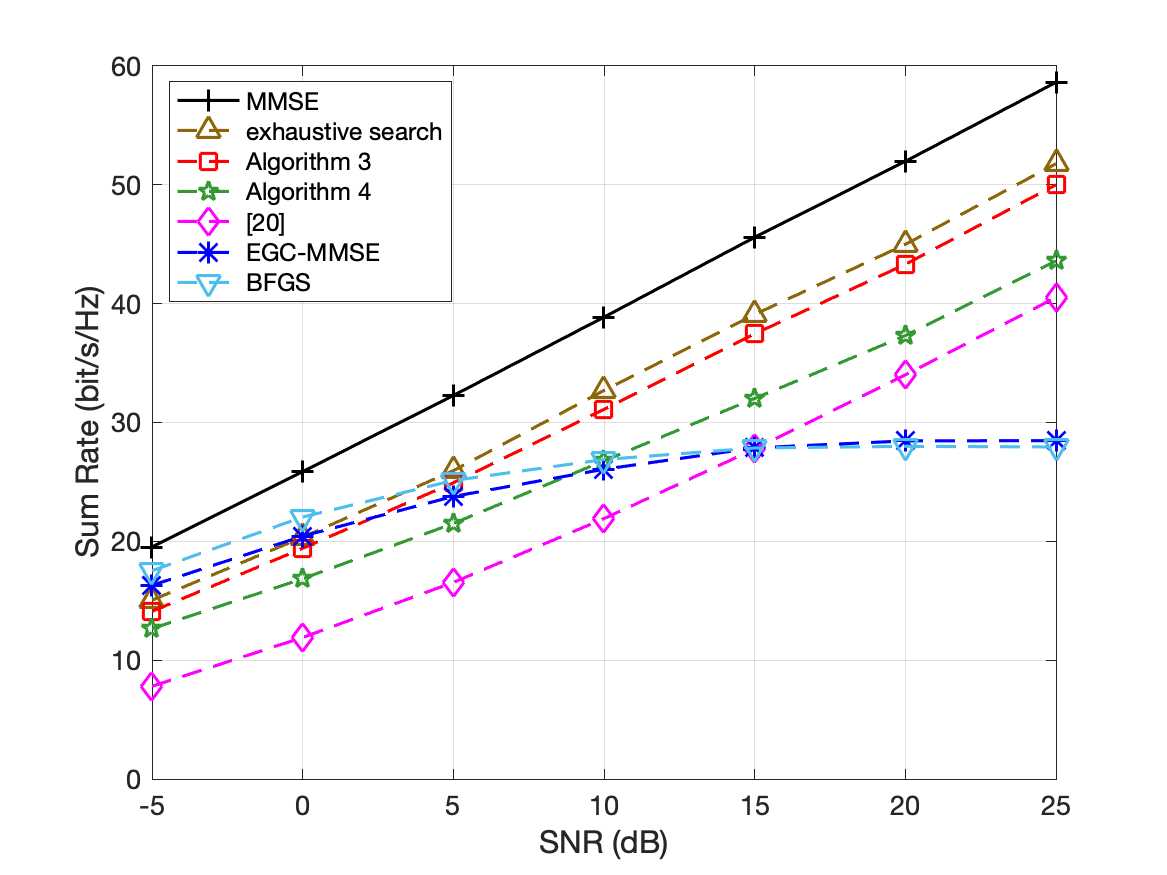}
	\caption{\footnotesize{Achievable sum rate versus Tx SNR (ISR = $0$ dB, $\Psi = 2$, $\Gamma_\psi = 1$,  $M = 8$, and $N = 16$).}} 
	\label{Fig_3}
\end{figure}

Figure~\ref{Fig_3} illustrates the achievable sum rate versus the SNR (i.e., ${P_{\rm U}}/{N_0}$), where $\Psi = 2$ inter-cell interferences with interference-to-signal ratio (ISR) fixed to $0$ dB (i.e., $P_{\rm I} = P_{\rm U}$) are accounted for. Setting the interference power equal to the signal power illustrates a substantial level of interference. It is seen that the sum rates of the optimal MMSE, the exhaustive search scheme, the proposed Algorithms~\ref{Algorithm-3} and \ref{Algorithm-4}, and the successive KHB \cite{Zhu2017} schemes increase linearly with SNR due to their excellence in interference mitigation. By contrast, the sum rates of BFGS and EGC-MMSE become saturated as SNR increases because of the Euclidean distance error to the MMSE beamformer for the first hybrid beamformer and the incapability of interference mitigation for the second. More specifically, the two proposed schemes outperform the successive KHB in \cite{Zhu2017}, and Algorithm~\ref{Algorithm-3} can achieve almost the same sum rate as the exhaustive search, with much lower computational complexity. The small sum-rate gap between Algorithm~\ref{Algorithm-3} and the exhaustive search scheme comes from the measure matrix \eqref{Eq_index} that is designed by equally allocating the channel coefficient to each Kronecker factor without considering the nonlinearity of unit-modulus factor multiplication.

\begin{figure}[!t]
	\centering
	\includegraphics [width = 0.475\textwidth,clip,keepaspectratio]{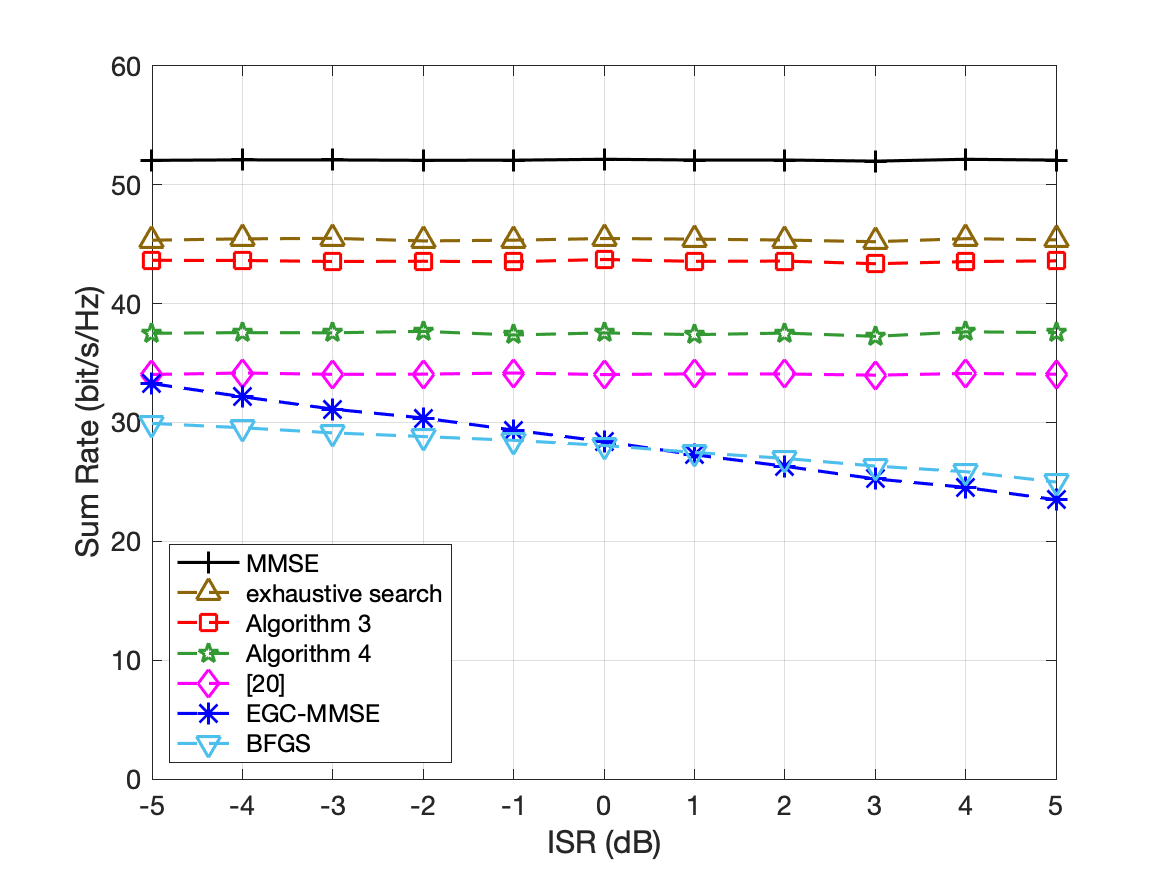}
	\caption{\footnotesize{Achievable sum rate versus Tx ISR (SNR = $20$~dB, $\Psi = 2$, $\Gamma_\psi = 1$, $M = 8$, and $N = 16$).}} 
	\label{Fig_4}
\end{figure}

\begin{figure}[!t]
	\centering
	\includegraphics [width = 0.475\textwidth,clip,keepaspectratio]{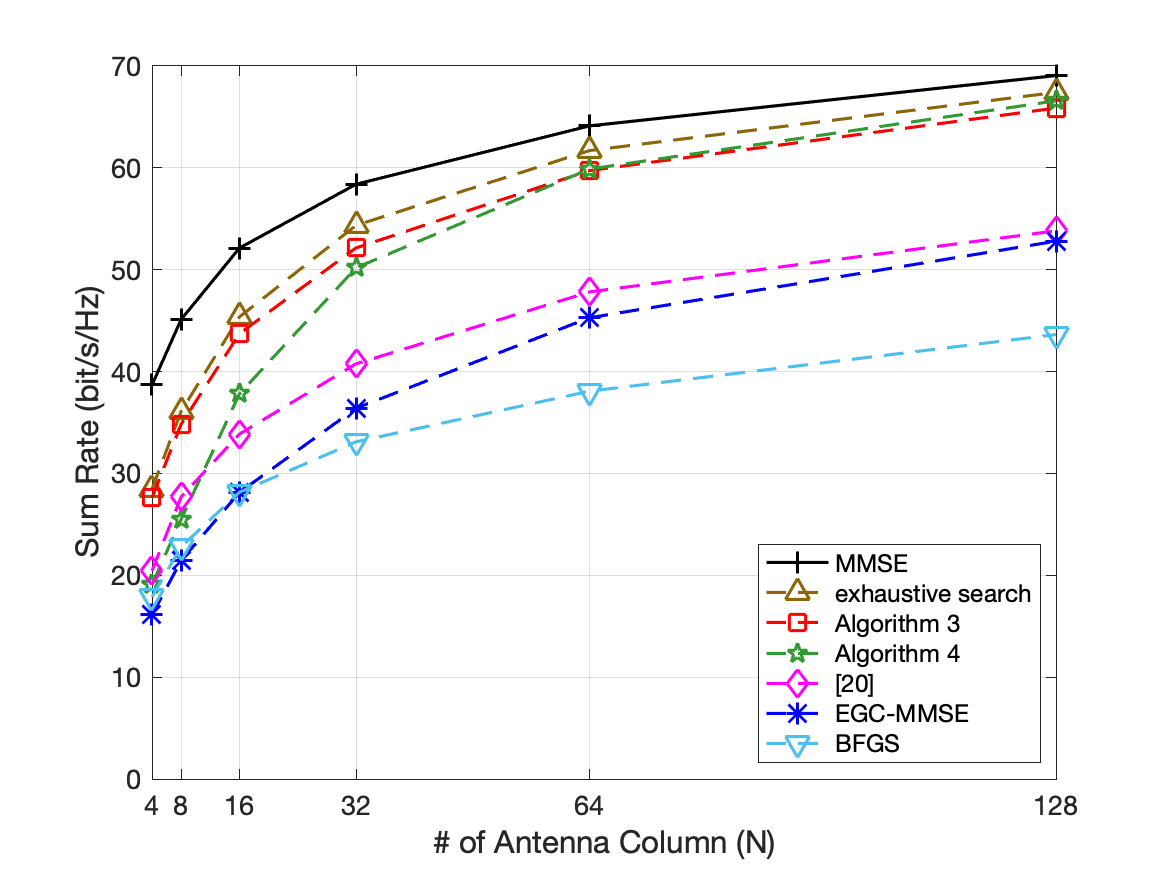}
	\caption{\footnotesize{Achievable sum rate versus column numbers of antenna array (SNR = $20$ dB, ISR = $0$ dB, $\Psi = 2$, $\Gamma_\psi = 1$, and $M = 8$).}} 
	\label{Fig_5}
\end{figure}

Figure~\ref{Fig_4} shows the achievable sum rate versus ISR in dB, where the SNR is fixed to $20$ dB, and there are $\Psi = 2$ inter-cell interferences. Setting the SNR to 20 dB reflects a typical value found in real-world scenarios. It is observed that the sum rate of the proposed Algorithm \ref{Algorithm-3} is about $85\%$ of that of the optimal MMSE scheme, over $95\%$ of that of the exhaustive search scheme, and $30\%$ higher than that of the successive KHB proposed in \cite{Zhu2017}. This significant performance gain comes from the dynamic factor allocation of Kronecker decomposition performed in Section~\ref{Subsection-KFA}. Also, the sum rate of the proposed Algorithm \ref{Algorithm-4} is higher than that of the successive KHB in \cite{Zhu2017}. Moreover, it is shown that the sum rates of these five schemes are robust to the strength of ISR because inter-cell interferences are eliminated. However, the sum rates of BFGS and EGC-MMSE schemes decrease with ISR due to the limited capability of interference mitigation of the BFGS scheme or the total incapability of interference mitigation of the EGC-MMSE scheme. 

Figure~\ref{Fig_5} shows the achievable sum rate versus the column number (i.e., $N$) of the UPA equipped at the BS, where the row number of the UPA is fixed to $M = 8$ and there are $\Psi = 2$ inter-cell interferences with ISR = $0$ dB. It is clear that, for any beamformer under study, the achievable sum rate increases logarithmically with $N$, as expected. The sum rate of Algorithm~\ref{Algorithm-3} is higher than that of the successive KHB in \cite{Zhu2017}, BFGS, and EGC-MMSE schemes and approaches that of the exhaustive search scheme under any antenna configuration. When the number of antennas is large enough (e.g., $N\geq 64$ in Fig.~\ref{Fig_5}), the proposed Algorithm \ref{Algorithm-4} outperforms Algorithm~\ref{Algorithm-3}. This observation accords well with the result reported in \cite{9505313}, where any beam misalignment degrades system performance significantly when an ultra-massive MIMO is concerned due to its highly directional beam. 

\begin{figure}[!t]
	\centering
	\includegraphics [width = 0.475\textwidth,clip,keepaspectratio]{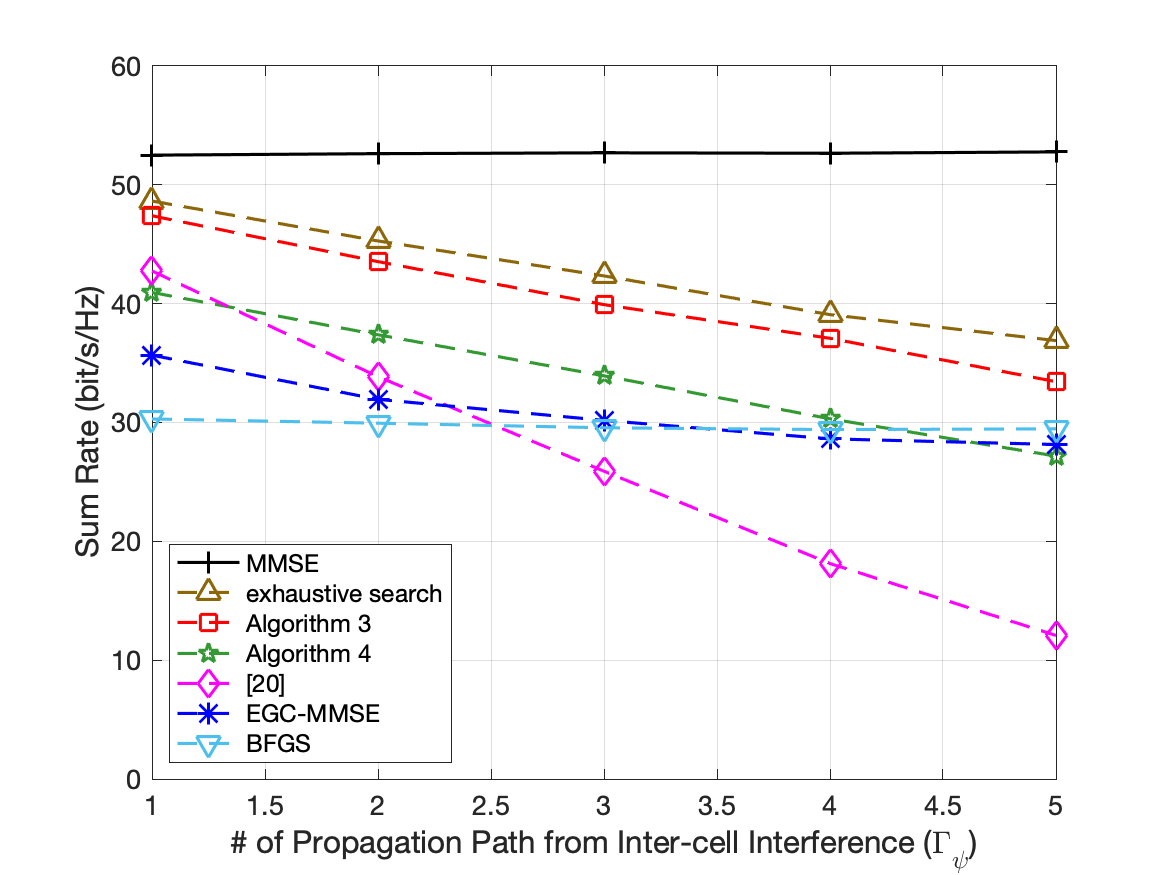}
	\caption{\footnotesize{Achievable sum rate versus the number of interference multi-paths (SNR = $20$ dB, ISR = $0$ dB, $\Psi = 1$,  $M = 8$, and $N = 16$).}} 
	\label{Fig_6}
\end{figure}

Figure~\ref{Fig_6} illustrates the achievable sum rate versus the number of interference multi-paths $\Gamma$, where a single inter-cell interference is considered with ISR fixed to $0$ dB. It is observed that the sum rates of the four KHB schemes decrease with the number of multi-paths, whereas the sum rates of the MMSE, BFGS, and EGC-MMSE schemes remain almost unvaried because of the fixed ISR. With the increasing number of interference propagation paths in the two proposed schemes, more Kronecker factors are consumed for null inter-cell interference components. Hence, the remaining DoFs for signal enhancement become fewer. Therefore, to make a tradeoff between interference cancellation and signal enhancement in real-world massive MIMO systems, the dominant interference components with relatively high power can be canceled out by Kronecker factors in the analog domain instead of nulling all interference components. In contrast, the others with relatively low power are treated as noise to release some DoFs of Kronecker factors for subsequent signal enhancement.

\section{Concluding Remarks}
\label{Section_Conclusion}
This paper designs a hybrid beamformer for the uplink transmission of a multi-cell mmWave FD-MIMO system using the primitive Kronecker decomposition and dynamic factor allocation strategies. In particular, the analog beamformer applies to null inter-cell interferences and enhances the desired signals. Then, a low-complexity hybrid beamformer is developed to slow the adjustment speed of the analog beamformer. Next,  we identified the optimal antenna configuration to exploit the spatial DoFs fully. Simulation results corroborated the two proposed beamformers' effectiveness and robustness against the strength of inter-cell interferences compared to the benchmark schemes. In real-world applications, the Kronecker factors of the beamforming vector can be efficiently allocated to suppress strong interferences instead of all inter-cell interferences, thus allowing a tradeoff between interference mitigation and signal enhancement.

\bibliographystyle{IEEEtran} 

\bibliography{References}

\begin{thebibliography}{10}
\providecommand{\url}[1]{#1}
\csname url@samestyle\endcsname
\providecommand{\newblock}{\relax}
\providecommand{\bibinfo}[2]{#2}
\providecommand{\BIBentrySTDinterwordspacing}{\spaceskip=0pt\relax}
\providecommand{\BIBentryALTinterwordstretchfactor}{4}
\providecommand{\BIBentryALTinterwordspacing}{\spaceskip=\fontdimen2\font plus
\BIBentryALTinterwordstretchfactor\fontdimen3\font minus
  \fontdimen4\font\relax}
\providecommand{\BIBforeignlanguage}[2]{{%
\expandafter\ifx\csname l@#1\endcsname\relax
\typeout{** WARNING: IEEEtran.bst: No hyphenation pattern has been}%
\typeout{** loaded for the language `#1'. Using the pattern for}%
\typeout{** the default language instead.}%
\else
\language=\csname l@#1\endcsname
\fi
#2}}
\providecommand{\BIBdecl}{\relax}
\BIBdecl

\bibitem{8626085}
E.~{Bj\"{o}rnson}, L.~{Van der Perre}, S.~{Buzzi}, and E.~G. {Larsson},
  ``Massive {MIMO} in sub-6 {GHz} and {mmWave}: Physical, practical, and
  use-case differences,'' \emph{IEEE Wireless Commun.}, vol.~26, no.~2, pp.
  100--108, Apr. 2019.

\bibitem{1203154}
A.~{Goldsmith}, S.~A. {Jafar}, N.~{Jindal}, and S.~{Vishwanath}, ``Capacity
  limits of {MIMO} channels,'' \emph{IEEE J. Sel. Areas Commun.}, vol.~21,
  no.~5, pp. 684--702, June 2003.

\bibitem{Ayach2014}
O.~E. Ayach, S.~Rajagopal, S.~Abu-Surra, Z.~Pi, and R.~W. Heath, ``Spatially
  sparse precoding in millimeter wave {MIMO} systems,'' \emph{IEEE Trans.
  Wireless Commun.}, vol.~13, no.~3, pp. 1499--1513, Mar. 2014.

\bibitem{Yu2016}
X.~{Yu}, J.~{Shen}, J.~{Zhang}, and K.~B. {Letaief}, ``Alternating minimization
  algorithms for hybrid precoding in millimeter wave {MIMO} systems,''
  \emph{IEEE J. Sel. Topics Signal Process.}, vol.~10, no.~3, pp. 485--500,
  Apr. 2016.

\bibitem{Jin2018}
J.~{Jin}, Y.~R. {Zheng}, W.~{Chen}, and C.~{Xiao}, ``Hybrid precoding for
  millimeter wave {MIMO} systems: A matrix factorization approach,'' \emph{IEEE
  Trans. Wireless Commun.}, vol.~17, no.~5, pp. 3327--3339, May 2018.

\bibitem{Peken2020}
T.~{Peken}, S.~{Adiga}, R.~{Tandon}, and T.~{Bose}, ``Deep learning for {SVD}
  and hybrid beamforming,'' \emph{IEEE Trans. Wireless Commun.}, vol.~19,
  no.~10, pp. 6621--6642, Oct. 2020.

\bibitem{Tse2005}
D.~Tse and P.~Viswanath, \emph{Fundamentals of Wireless Communication}.\hskip
  1em plus 0.5em minus 0.4em\relax Cambridge University Press, 2005.

\bibitem{Li2017}
A.~{Li} and C.~{Masouros}, ``Hybrid precoding and combining design for
  millimeter-wave multi-user {MIMO} based on {SVD},'' in \emph{Proc. IEEE Int.
  Conf. Commun. (ICC)}, May 2017, pp. 1--6.

\bibitem{Nguyen2017}
D.~H.~N. {Nguyen}, L.~B. {Le}, T.~{Le-Ngoc}, and R.~W. {Heath}, ``Hybrid {MMSE}
  precoding and combining designs for {mmWave} multiuser systems,'' \emph{IEEE
  Access}, vol.~5, pp. 19\,167--19\,181, Sept. 2017.

\bibitem{Shi2018}
Q.~{Shi} and M.~{Hong}, ``Spectral efficiency optimization for millimeter wave
  multiuser {MIMO} systems,'' \emph{IEEE J. Sel. Topics Signal Process.},
  vol.~12, no.~3, pp. 455--468, Apr. 2018.

\bibitem{Alkhateeb2015}
A.~Alkhateeb, G.~Leus, and R.~Heath, ``Limited feedback hybrid precoding for
  multi-user millimeter wave systems,'' \emph{IEEE Trans. Wireless Commun.},
  vol.~14, no.~11, pp. 6481--6494, Nov. 2015.

\bibitem{Zhang2020}
Y.~{Zhang}, J.~{Du}, Y.~{Chen}, X.~{Li}, K.~M. {Rabie}, and R.~{Kharel},
  ``Near-optimal design for hybrid beamforming in {mmWave} massive multi-user
  {MIMO} systems,'' \emph{IEEE Access}, vol.~8, pp. 129\,153--129\,168, July
  2020.

\bibitem{9419762}
G.~M. Zilli and W.-P. Zhu, ``Constrained tensor decomposition-based hybrid
  beamforming for mmwave massive {MIMO-OFDM} communication systems,''
  \emph{IEEE Trans. Veh. Technol.}, vol.~70, no.~6, pp. 5775--5788, Apr. 2021.

\bibitem{10355874}
B.~Lin, A.~Liu, M.~Lei, and H.~Zhou, ``Low-complexity two-timescale hybrid
  precoding for {mmWave} massive {MIMO}: A group-and-codebook based approach,''
  \emph{IEEE Trans. Wireless Commun.}, vol.~23, no.~7, pp. 7263--7277, July
  2024.

\bibitem{Liu2019}
J.~{Liu}, M.~{Sheng}, R.~{Lyu}, and J.~{Li}, ``Performance analysis and
  optimization of {UAV} integrated terrestrial cellular network,'' \emph{IEEE
  Internet Things J.}, vol.~6, no.~2, pp. 1841--1855, Apr. 2019.

\bibitem{10381632}
X.~Fan, P.~Wu, and M.~Xia, ``Air-to-ground communications beyond {5G}: {UAV}
  swarm formation control and tracking,'' \emph{IEEE Trans. Wireless Commun.},
  vol.~23, no.~7, pp. 8029--8043, July 2024.

\bibitem{8976426}
Y.~{Li}, M.~{Xia}, and S.~{A\"{i}ssa}, ``Coordinated multi-point transmission:
  A {Poisson-Delaunay} triangulation based approach,'' \emph{IEEE Trans.
  Wireless Commun.}, vol.~19, no.~5, pp. 2946--2959, May 2020.

\bibitem{Nasir2020}
A.~A. {Nasir}, H.~D. {Tuan}, T.~Q. {Duong}, H.~V. {Poor}, and L.~{Hanzo},
  ``Hybrid beamforming for multi-user millimeter-wave networks,'' \emph{IEEE
  Trans. Veh. Technol.}, vol.~69, no.~3, pp. 2943--2956, Mar. 2020.

\bibitem{9349195}
J.~Zhan and X.~Dong, ``Interference cancellation aided hybrid beamforming for
  mmwave multi-user massive {MIMO} systems,'' \emph{IEEE Trans. Veh. Technol.},
  vol.~70, no.~3, pp. 2322--2336, Feb. 2021.

\bibitem{Zhu2017}
G.~Zhu, K.~Huang, V.~K.~N. Lau, B.~Xia, X.~Li, and S.~Zhang, ``Hybrid
  beamforming via the {Kronecker} decomposition for the millimeter-wave massive
  {MIMO} systems,'' \emph{IEEE J. Sel. Areas Commun.}, vol.~35, no.~9, pp.
  2097--2114, Sept. 2017.

\bibitem{Xu2017}
G.~{Xu}, Y.~{Li}, J.~{Yuan}, R.~{Monroe}, S.~{Rajagopal}, S.~{Ramakrishna},
  Y.~H. {Nam}, J.~{Seol}, J.~{Kim}, M.~M.~U. {Gul}, A.~{Aziz}, and J.~{Zhang},
  ``Full dimension {MIMO} ({FD-MIMO}): Demonstrating commercial feasibility,''
  \emph{IEEE J. Sel. Areas Commun.}, vol.~35, no.~8, pp. 1876--1886, Aug. 2017.

\bibitem{7060514}
B.~{Mondal}, T.~A. {Thomas}, E.~{Visotsky}, F.~W. {Vook}, A.~{Ghosh}, Y.~{Nam},
  Y.~{Li}, J.~{Zhang}, M.~{Zhang}, Q.~{Luo}, Y.~{Kakishima}, and K.~{Kitao},
  ``{3D} channel model in {3GPP},'' \emph{IEEE Commun. Mag.}, vol.~53, no.~3,
  pp. 16--23, 2015.

\bibitem{Hardy1938}
G.~H. Hardy, E.~M. Wright, and A.~Wiles, \emph{An Introduction to the Theory of
  Numbers}, 6th~ed.\hskip 1em plus 0.5em minus 0.4em\relax Oxford University
  Press, 2008.

\bibitem{38.901}
2019, 3GPP, TR 38.901: ``Study on channel model for frequencies from 0.5 to 100
  GHz (Release 16)".

\bibitem{10214216}
H.~Huang, K.~Wang, P.~Wu, J.~Zhang, and M.~Xia, ``Flexible {3GPP} {MIMO}
  channel modeling and calibration with spatial consistency,'' \emph{IEEE
  Access}, vol.~11, pp. 85\,137--85\,154, Aug. 2023.

\bibitem{9217058}
C.~Chen, Y.~Wang, S.~Aïssa, and M.~Xia, ``Low-complexity hybrid analog and
  digital precoding for mmwave mimo systems,'' in \emph{2020 IEEE 31st Annual
  International Symposium on Personal, Indoor and Mobile Radio Communications},
  2020, pp. 1--6.

\bibitem{Graham2018}
A.~Graham, \emph{Kronecker Products and Matrix Calculus with
  Applications}.\hskip 1em plus 0.5em minus 0.4em\relax Dover Publications,
  2018.

\bibitem{Bjornson2017}
E.~Bj\"{o}rnson, J.~Hoydis, and L.~Sanguinetti, \emph{Massive {MIMO} Networks:
  Spectral, Energy, and Hardware Efficiency}.\hskip 1em plus 0.5em minus
  0.4em\relax Foundations and Trends in Signal Processing, 2017, vol.~11.

\bibitem{Alhujaili2019}
K.~{Alhujaili}, V.~{Monga}, and M.~{Rangaswamy}, ``Transmit {MIMO} radar
  beampattern design via optimization on the complex circle manifold,''
  \emph{IEEE Trans. Signal Process.}, vol.~67, no.~13, pp. 3561--3575, July
  2019.

\bibitem{Bernstein2009}
D.~S. Bernstein, \emph{Matrix Mathematics: Theory, Facts, and Formulas},
  2nd~ed.\hskip 1em plus 0.5em minus 0.4em\relax Princeton University Press,
  2009.

\bibitem{Verdu1998}
S.~Verd\'{u}, \emph{Multiuser Detection}.\hskip 1em plus 0.5em minus
  0.4em\relax Cambridge University Press, 1998.

\bibitem{Fan2018}
D.~{Fan}, F.~{Gao}, Y.~{Liu}, Y.~{Deng}, G.~{Wang}, Z.~{Zhong}, and
  A.~{Nallanathan}, ``Angle domain channel estimation in hybrid millimeter wave
  massive {MIMO} systems,'' \emph{IEEE Trans. Wireless Commun.}, vol.~17,
  no.~12, pp. 8165--8179, Dec. 2018.

\bibitem{9505313}
Y.~Chen, L.~Yan, C.~Han, and M.~Tao, ``Millidegree-level direction-of-arrival
  estimation and tracking for terahertz ultra-massive {MIMO} systems,''
  \emph{IEEE Trans. Wireless Commun.}, vol.~21, no.~2, pp. 869--883, Feb. 2022.

\end{thebibliography}
\vfill
\end{document}